\documentclass[acmtog]{acmart}
\acmSubmissionID{1881}

\usepackage{booktabs} 

\usepackage{multirow}

\usepackage[ruled]{algorithm2e} 

\SetAlFnt{\small}
\SetAlCapFnt{\small}
\SetAlCapNameFnt{\small}
\SetAlCapHSkip{0pt}

\acmJournal{TOG}
\newcommand{\revise}[1]{{#1}}

\newcommand{\name}{Fuse3D}

\newcommand{\refFig}[1]{Figure \ref{#1}}

\newcommand{\refSec}[1]{Section \ref{#1}}
\newcommand{\refTab}[1]{Table \ref{#1}}

\copyrightyear{2025}
\acmYear{2025}
\setcopyright{acmlicensed}\acmConference[SA Conference Papers '25]{SIGGRAPH Asia 2025 Conference Papers}{December 15--18, 2025}{Hong Kong, Hong Kong}
\acmBooktitle{SIGGRAPH Asia 2025 Conference Papers (SA Conference Papers '25), December 15--18, 2025, Hong Kong, Hong Kong}
\acmDOI{10.1145/3757377.3763943}
\acmISBN{979-8-4007-2137-3/2025/12}

\begin{document}
\title{Fuse3D: Generating 3D Assets Controlled by Multi-Image Fusion}
 
\author{Xuancheng Jin}
\affiliation{
  \institution{State Key Laboratory of CAD\&CG, Zhejiang University, Zhejiang University of Technology}
  \city{Hangzhou}
  \country{China}
}
\authornote{Xuancheng Jin conducted this work during his internship at Zhejiang University.}
\authornote{Equal Contribution.}

\author{Rengan Xie}
\affiliation{
  \institution{State Key Laboratory of CAD\&CG, Zhejiang University}
  \city{Hangzhou}
  \country{China}
}
\authornotemark[2]

\author{Wenting Zheng}
\affiliation{
  \institution{State Key Laboratory of CAD\&CG, Zhejiang University}
  \city{Hangzhou}
  \country{China}
}

\author{Rui Wang}
\affiliation{
  \institution{State Key Laboratory of CAD\&CG, Zhejiang University}
  \city{Hangzhou}
  \country{China}
}

\author{Hujun Bao}
\affiliation{
  \institution{State Key Laboratory of CAD\&CG, Zhejiang University}
  \city{Hangzhou}
  \country{China}
}

\author{Yuchi Huo}
\affiliation{
  \institution{State Key Laboratory of CAD\&CG, Zhejiang University}
  \city{Hangzhou}
  \country{China}
}
\authornote{Corresponding Author.}

\begin{teaserfigure}
  \centering
    \includegraphics[width=1.0\linewidth]{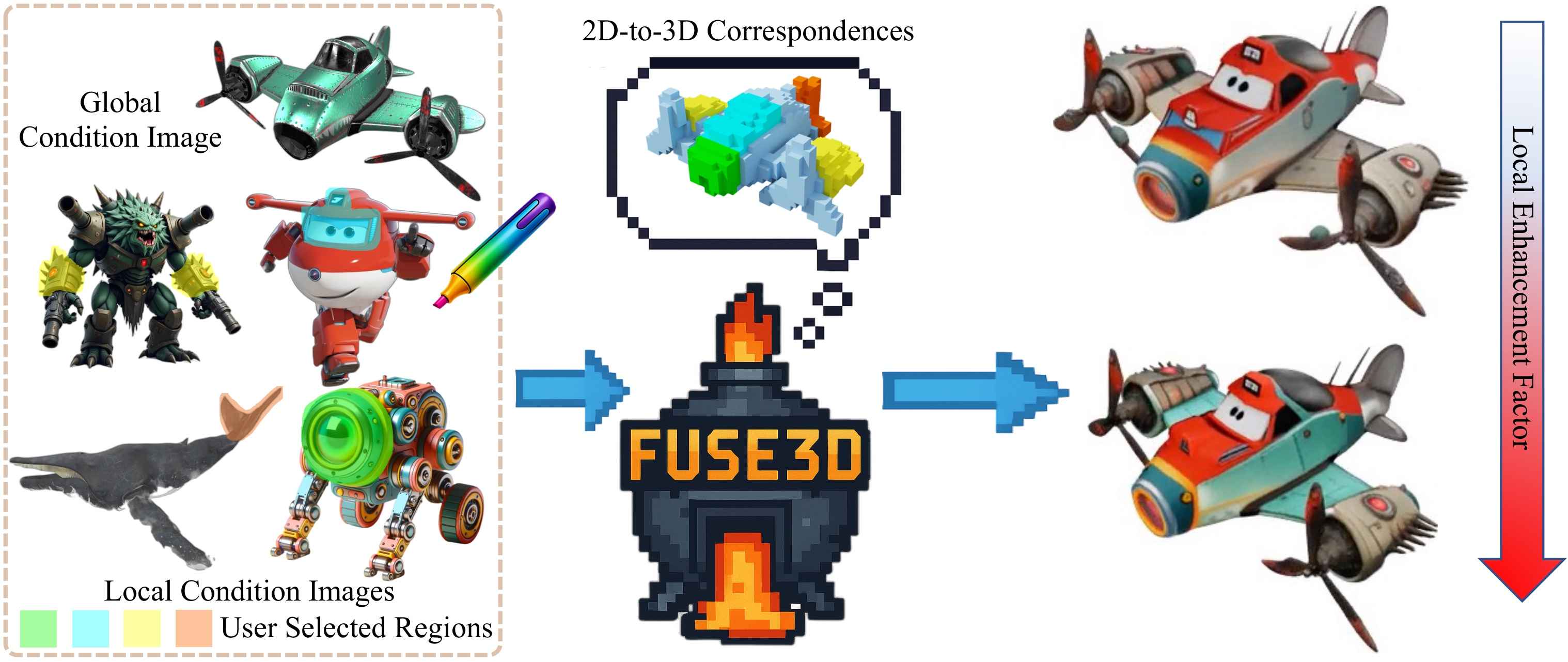}
    \caption{
    \name~enables generating controllable 3D assets from multiple condition images. Furthermore, users can specify particular regions in the condition images to guide the synthesis of semantically and geometrically aligned areas in the generated 3D asset.
    }
    \label{fig:teaser}
\end{teaserfigure}

\begin{abstract}

Recently, generating 3D assets with the control of condition images has achieved impressive quality. 
However, existing 3D generation methods are limited to handling a single control objective and lack the ability to utilize multiple images to independently control different regions of a 3D asset, which hinders their flexibility in applications.
We propose \name, a novel method that enables generating 3D assets under the control of multiple images, allowing for the seamless fusion of multi-level regional controls from global views to intricate local details.
First, we introduce a Multi-Condition Fusion Module to integrate the visual features from multiple image regions.
Then, we propose a method to automatically align user-selected 2D image regions with their associated 3D regions based on semantic cues.
Finally, to resolve control conflicts and enhance local control features from multi-condition images, we introduce a Local Attention Enhancement Strategy that flexibly balances region-specific feature fusion.
Overall, we introduce the first method capable of controllable 3D asset generation from multiple condition images.
The experimental results indicate that \name~can flexibly fuse multiple 2D image regions into coherent 3D structures, resulting in high-quality 3D assets.
\revise{Code and data for this paper are at https://jinnmnm.github.io/Fuse3d.github.io/.}


\end{abstract}



\maketitle

\section{Introduction}
Acquiring realistic 3D content efficiently and accurately remains a fundamental challenge for various applications, such as film production, gaming, mixed reality, and industrial design. \revise{Recently, 3D generation models \cite{trellis,clay} have shown great potential for generating 3D assets from input texts and images, bypassing tedious manual modeling by artists.} However, existing 3D generation methods often have single-focused objectives, aiming to generate a 3D asset that globally aligns with the input image and text, making it challenging to achieve fine-grained, part-level control. 

Although 3D generation models driven by text input can enhance local control over the generated object by incorporating detailed text descriptions of specific regions, text-based control often lacks precision and tends to be ambiguous. We aim to propose a method that can utilize control features from multiple condition images to generate 3D assets that are controllable for specific regions. 

In terms of controllability, 2D image generation is ahead of 3D generation. Ctrl-X\cite{ctrlx} and ControlNet\cite{controlNet} enhance the capability of Stable Diffusion\cite{stable_diffusion}  through advanced feature composition and engineering, enabling precise control over the final 2D image using multiple conditions. These methods reveal that generative models trained on extensive datasets possess a deeper understanding of the 3D world, with their outputs primarily constrained by input guidance rather than by the intrinsic potential of the models.
Inspired by the success of controllable 2D image generation, we propose \name~to effectively fuse control features from multiple condition images and effectively inject them into the 3D generation process. 3D generation involves complex target structures, making it inherently challenging.

First, given multiple images containing user-selected 2D regions for different control objectives, we divide these images into a global condition image and local condition images based on the area of the condition regions. The key challenge is to effectively fuse them into a unified conditional representation that maintains their spatial and semantic integrity. To tackle this, we propose a Multi-Condition Fusion Module (MCFM) that aggregates the control features from condition images. Multiple condition images are fed into the DINOv2~\cite{dinov2} to extract feature tokens containing semantic and spatial information in selected 2D regions, which are then deliberately fused under the guidance of 2D region masks to produce unified condition tokens encapsulating multiple control objectives.

In Fused3D, we choose TRELLIS\cite{trellis}, a powerful 3D-native generation model, as our main generation backbone, which can generate structured latent representations from input condition and further decode them into 3D structures. As our unified condition tokens and the condition space of TRELLIS both lie in the feature space of DINOv2, establishing a correspondence between the condition tokens and the structured latent representations enables precise local control over the 3D structure. This introduces the second challenge, which involves establishing precise spatial correspondences between the selected 2D tokens and their target 3D regions. We observe that the cross-attention layer in the TRELLIS model can capture the association strength between condition tokens and 3D structure voxels, which is positively correlated with their attention values. Inspired by this observation, we derive a coarse voxel representation of the target 3D object from the global image using the Flow Transformer of TRELLIS and propose a 3D Semantic-Aware Alignment Strategy to establish accurate correspondences between condition tokens and 3D voxels. Specifically, this strategy encompasses bidirectional alignment. The forward alignment retrieves corresponding 3D voxels for image tokens based on local condition images, while the reverse alignment retrieves image tokens from the global condition image for 3D voxels that are not associated with any selected condition tokens.

Finally, despite being guided by the unified condition tokens, the generated structured latent may suffer from undesired blending, as control features from different regions may overlap and interfere within shared 3D regions. To mitigate this, we introduce a Local Attention Enhancement Strategy that leverages the 2D–to-3D correspondences obtained in the forward alignment to apply region-aware attention scaling. This mechanism allows flexible control over the fusion strength, enhancing the 3D local features corresponding to 2D regions while ensuring the smooth blending of multiple compositions.
After generating the final structured latent, the pretrained VAE decoder from TRELLIS transforms the it into a full 3D asset. 

Overall, the main contributions of this work can be summarized as follows.
\begin{itemize}
\item To the best of our knowledge, \name~is the first method that utilizes control objectives derived from multiple 2D condition images to precisely guide the generation of target 3D assets, from global to local 3D structures.
\item We propose a Multi-Condition Fusion Module (MCFM) that aggregates the control features from condition images. 
\item We propose a 3D Semantic-Aware Alignment Strategy to automatically establish correspondence between 2D condition tokens and 3D voxels.
\item We propose a Local Attention Enhancement strategy that addresses the conflicts among multiple control objectives, enabling controllable 3D local feature enhancement and smooth fusion of different controlled components.
\end{itemize}

\section{related work}

\subsection{2D image Generation}

GANs\cite{GAN} have achieved remarkable success in generating impressive images across domains such as faces, scenes, and artworks. StyleGAN\cite{styleGAN} further improved image quality and diversity by introducing hierarchical latent control and style modulation. However, challenges like mode collapse limit the scalability of GAN. Diffusion models\cite{ddim, ddpm} address these issues by modeling generation as a gradual denoising process, generating structured images with high quality and diversity.

To control image generation more effectively, early methods adopted classifier guidance~\cite{classifierGuidance}, which guides diffusion sampling using gradients from a pretrained image classifier. However, the fixed classifier may limit expressiveness. Later, classifier-free guidance (CFG)~\cite{cfg} emerged as a widely adopted alternative, enabling conditional sampling (e.g., with text prompts) by training the diffusion model with and without conditioning inputs, and dynamically blending the outputs during inference.
With CFG as the backbone, recent advances in text-to-image generation—such as DALL·E~\cite{dalle2}, Imagen~\cite{imagen}, and Stable Diffusion~\cite{stable_diffusion}—enable controllable high-resolution image synthesis from textual prompts. 
\revise{In parallel, tuning-free 2D editing methods such as Prompt-to-Prompt~\cite{prompt2prompt} and MasaCtrl~\cite{masactrl} have demonstrated fine-grained control by modulating attention maps or aligning tokens without retraining. 
These approaches highlight the effectiveness of attention modulation for precise regional control in 2D.} 

To further explore the potential of 2D image generation models, structure-conditioned diffusion models have emerged alongside text-based prompts, offering spatially precise control signals.
Notably, image-to-image frameworks~\cite{controlNet,ctrlx} inject guidance from structured modalities (e.g., depth, canny, or pose) into the diffusion process, leveraging carefully designed feature blending in network. 
IPAdapter~\cite{ipAdapter} introduces an additional token adapter that transforms the condition image into feature tokens, which are subsequently fused with the feature tokens extracted from the text prompt. 
The success in controllable 2D image synthesis with complex visual attributes inspires us to construct a framework that better blends control features from multiple condition images to achieve precise generation of the target 3D structure.
\vspace{-0.8em}


\subsection{3D Generative Models}

Generative modeling of 3D objects has rapidly evolved, particularly with the advent of diffusion models, which have demonstrated remarkable success in generating diverse 3D representations such as point clouds~\cite{pointE}, implicit fields~\cite{SDFusion, shapeE, diffusionSDF}, explicit meshes~\cite{GET3D}, \revise{and multi-view images~\cite{zero123,wonder3D,edify3d}}. These methods directly model the probability distribution of 3D shapes, leveraging diffusion processes to achieve high-quality and diverse outputs.

More recently, approaches like DreamFusion~\cite{dreamFusion} and its follow-ups~\cite{magic3D, fantasia3D, csd, prolificDream} introduced the Score Distillation Sampling (SDS) loss to distill 2D generative priors into a 3D reconstruction pipeline. \revise{Extensions of this line of work also target specific domains, such as human-centric generation~\cite{arc2avatar,humanNorm}.} Despite achieving promising results, these methods typically require significant computation and face inherent instability issues, such as the Janus problem.

To mitigate these challenges, recent advances like LRM~\cite{lrm,lgm,ldm} have pursued end-to-end models trained directly on 3D data, eliminating reliance on noisy gradient signals from 2D priors. These methods use transformer-based architectures to predict neural fields or meshes directly from input conditions, significantly improving quality and reducing computational overhead.

Most recently, 3D native generative methods such as CLAY~\cite{clay} and TRELLIS~\cite{trellis} adopt a structured two-stage generation approach, leveraging VAE models to encode the target 3D representation into the latent space, followed by diffusion-based generation within the latent space. This decoupling has proven highly effective, yielding consistent and high-fidelity 3D outputs while enabling finer control over the generation process. In this work, we adopt TRELLIS as the backbone of the generation process due to its fully open-source nature.
\vspace{-0.8em}

\subsection{Controllable 3D Generation}

Controlling the structure and appearance of generated 3D content is crucial for applications such as interactive modeling and content creation.
Text-to-3D methods like DreamFusion~\cite{dreamFusion} and Magic3D~\cite{magic3D} provide high-level semantic control, but lack spatial specificity. 

Beyond text prompts, recent works have introduced many flexible inputs to control the generation, including sketch image inputs~\cite{control3D}, coarse geometry~\cite{coin3d, fantasia3D, latentNerf, skdream}, \revise{overall layout~\cite{cc3d}, and appearance image~\cite{ipdreamer}}.
DreamBooth3D~\cite{dreamBooth3D} relaxes the constraints on input condition images, extending its applicability beyond highly consistent multi-view images~\cite{instantMesh} to just a few casual images of a subject. However, these methods are still limited to global control over the generation of 3D objects, with little focus on precise local control.

Recent methods have started focusing on local control of generated objects, which mostly rely on local inpainting and editing approaches. FocalDreamer~\cite{focalDream} distills 2D edits into the 3D space through SDS-based optimization to achieve the fusion of two 3D structures. Progressive3D~\cite{progressive3D} decomposes a complete target 3D structure into multiple distinct local regions using 3D masks, and inpaints the target regions through iterative SDS-based distillation. A3D~\cite{A3D} generates a set of objects with similar 3D structures through iterative SDS-based distillation and achieves the fusion of multiple local 3D structures by manually selecting 3D regions for combination. These methods rely on iterative SDS optimization, making them time-intensive. 

MaskedLRM~\cite{meshEditing} proposes a specialized training strategy refine LRM model~\cite{lrm}, enabling local editing from a single-view image with only a single forward inference. \revise{Similarly, EditP23~\cite{editp23} introduces an edit-aware denoising scheme that propagates target-view edits across other views to achieve consistent 3D editing.}
More recent efforts SDFusion~\cite{SDFusion} explore integrating multiple input conditions, such as partial point clouds, images, and text, into the denoising process of signed distance function (SDF) diffusion models. However, it focuses solely on generating the SDF of target object SDF without incorporating appearance information and relies on partial shapes as input and cannot handle control from multiple conditional images. 
\revise{PartCrafter~\cite{partcrafter} trains a part-aware 3D DiT that generates per-part latents conditioned by one image with local/global attention, which are decoded into separate meshes. This decompositional design enables per-part generation, but it is not designed for multi-image, region-level fusion.}
In this work, we propose a method that fuses control features from multiple conditional images to regulate 3D structure generation, ranging from global shape to local details, and the entire generation process can be completed within 20 seconds.

\begin{figure*}
    \centering
    \includegraphics[width=\textwidth]{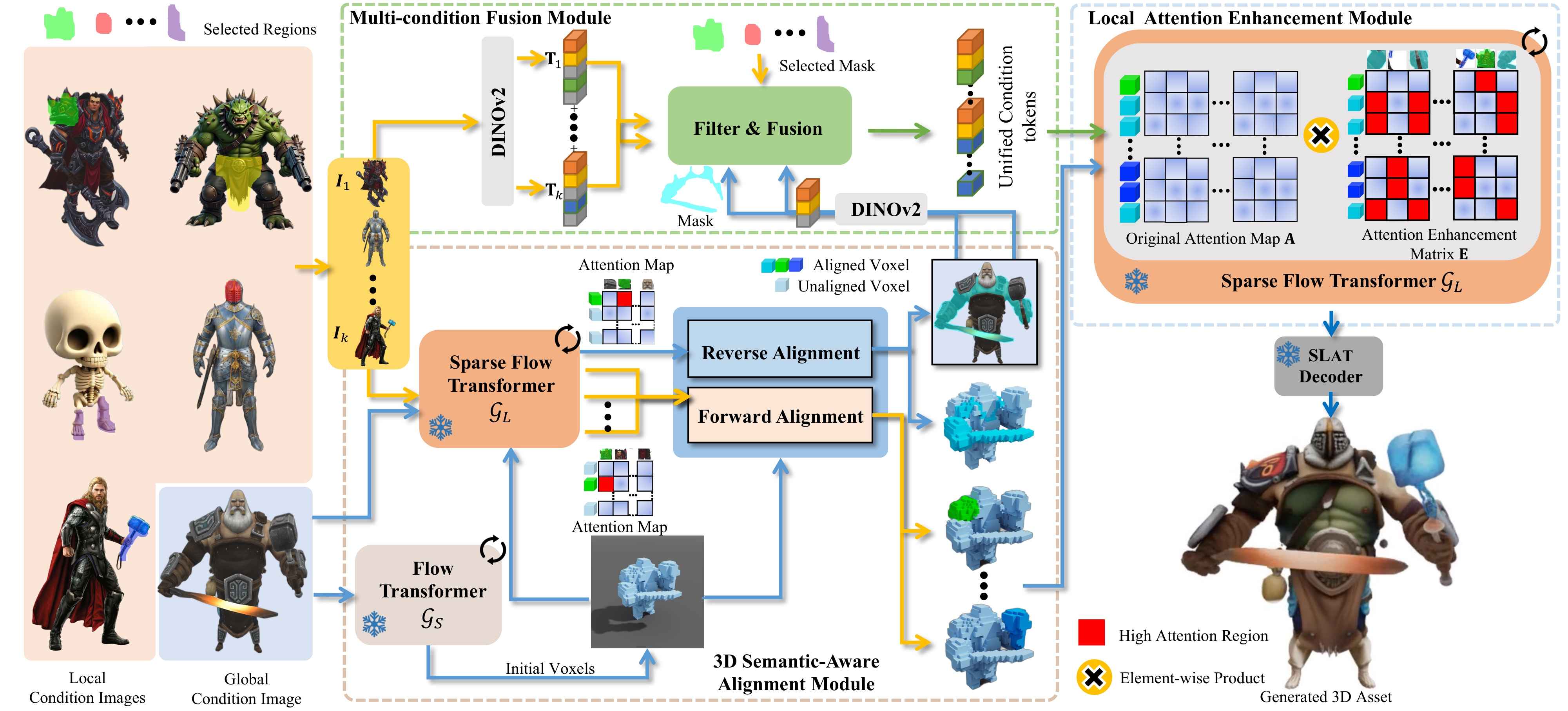}
    \vspace{-0.4cm}
    \caption{{\bf Overview. } 
        Given multiple condition images, we first utilize a Multi-Condition Fusion Module (MCFM) to obtain fused unified condition tokens. Subsequently, a 3D Semantic-Aware Alignment Module is used to establish the correspondence between the condition tokens and their target 3D structures. Finally, a Local Attention Enhancement Module is introduced to generate the target fused 3D asset.
    }
   \label{fig:overview}
   \vspace{-0.3cm}
\end{figure*}

\section{method}


We propose \name, a novel method that generates high-quality 3D assets under the guidance of multiple condition images, where each image controls specific features of the target 3D object, ranging from global to local. \refFig{fig:overview} shows a overview of our method. We briefly introduce TRELLIS in~\refSec{sec:prelim} and outline the preparatory steps before generating the fused 3D assets. Then, we describe in~\refSec{sec:MCFM} how to extract fused control tokens from multiple condition images. Subsequently, we introduce a 3D Semantic-Aware Alignment strategy in Sec 3.3 to establish the connection between condition tokens and the target 3D structure. Finally, we describe how to perform Local Attention Enhancement to obtain a target 3D asset with prominent local features in ~\refSec{Sec:localEnhance}.

\subsection{Preliminaries and Latent Initialization}\label{sec:prelim}
\paragraph{TRELLIS}
We adopt the \textsc{SLat} representation proposed in TRELLIS~\cite{trellis} as 3D latent space for our framework.
For a 3D asset $\mathcal O$, the \textsc{SLat} representation is defined as
\begin{equation}
    \boldsymbol{Z} = \{(\boldsymbol{z}_i,\boldsymbol{p}_i)\}_{i=1}^{L},\quad \boldsymbol{z}_i\in\mathbb{R}^C, \ \boldsymbol{p}_i\in \{0, 1,\ldots, N-1\} ^3, 
    \label{eq:slate}
\end{equation}
where $\boldsymbol{p}_i$ denotes the index of an active voxel in 3D grid, $\boldsymbol{z}_i$ is a local latent attached to voxel $\boldsymbol{p}_i$, $L$ is the total number of active voxels, $C$ is the channel dimension and $N$ is the spatial length of 3D grid.
Specifically, TRELLIS generates the 3D asset in a two-stage pipeline.
In the first stage, it generates the sparse structure $\{\boldsymbol{p}_i\}_{i=1}^{L}$ of $\mathcal{O}$.
This is achieved by using a transformer-based flow model~\cite{rectifiedFlow} $\boldsymbol{\mathcal{G}}_{\mathrm{S}}$ to produce a low-resolution feature grid $\boldsymbol{S}\in\mathbb{R}^{D\times D\times D \times C_\mathrm{S}}$, which is then decoded by a latent feature decoder $\mathcal{D}_S$ to obtain a dense binary 3D grid $\boldsymbol{O} \in \{0,1\}^{N\times N\times N}$.
The grid $\boldsymbol{O}$ is then converted to the positional index of active voxels $\{\boldsymbol{p}_i\}_{i=1}^{L}$.
In the second stage, TRELLIS uses another transformer-based flow model $\boldsymbol{\mathcal{G}}_{\mathrm{L}}$ to generate the corresponding latent features $\{\boldsymbol{z}_i\}_{i=1}^{L}$ for these active voxels.
The complete \textsc{SLat} representation $\boldsymbol{Z}$ is then processed through specialized decoders ($\mathcal{D}_{\mathrm{NeRF}}$, $\mathcal{D}_{\mathrm{Mesh}}$, or $\mathcal{D}_{\mathrm{GS}}$) to produce the final 3D asset $\mathcal{O}$ in various formats (NeRF, meshes, or 3DGS).

\paragraph{Voxel initialization from a global image.}
Given a set of condition images, we assign the image with the most highlighted control regions as the global condition image $I_{\text{g}}$, aimed at ensuring the global features of the generated 3D assets. The other images serve as local condition images, offering precise local control over the 3D assets. We extract an initial voxels
$\{\boldsymbol{p}_i^0\}_{i=1}^{L}$
using the pretrained transformer-based flow model $\boldsymbol{\mathcal{G}}_{\mathrm{S}}$.
These initial voxels $\{\boldsymbol{p}_i^0\}_{i=1}^{L}$ provides the basis for all subsequent generation performed directly in the \textsc{SLat} latent space.

\subsection{Multi-condition Fusion Module}\label{sec:MCFM}

\begin{figure}
    \centering
    \includegraphics[width=\linewidth]{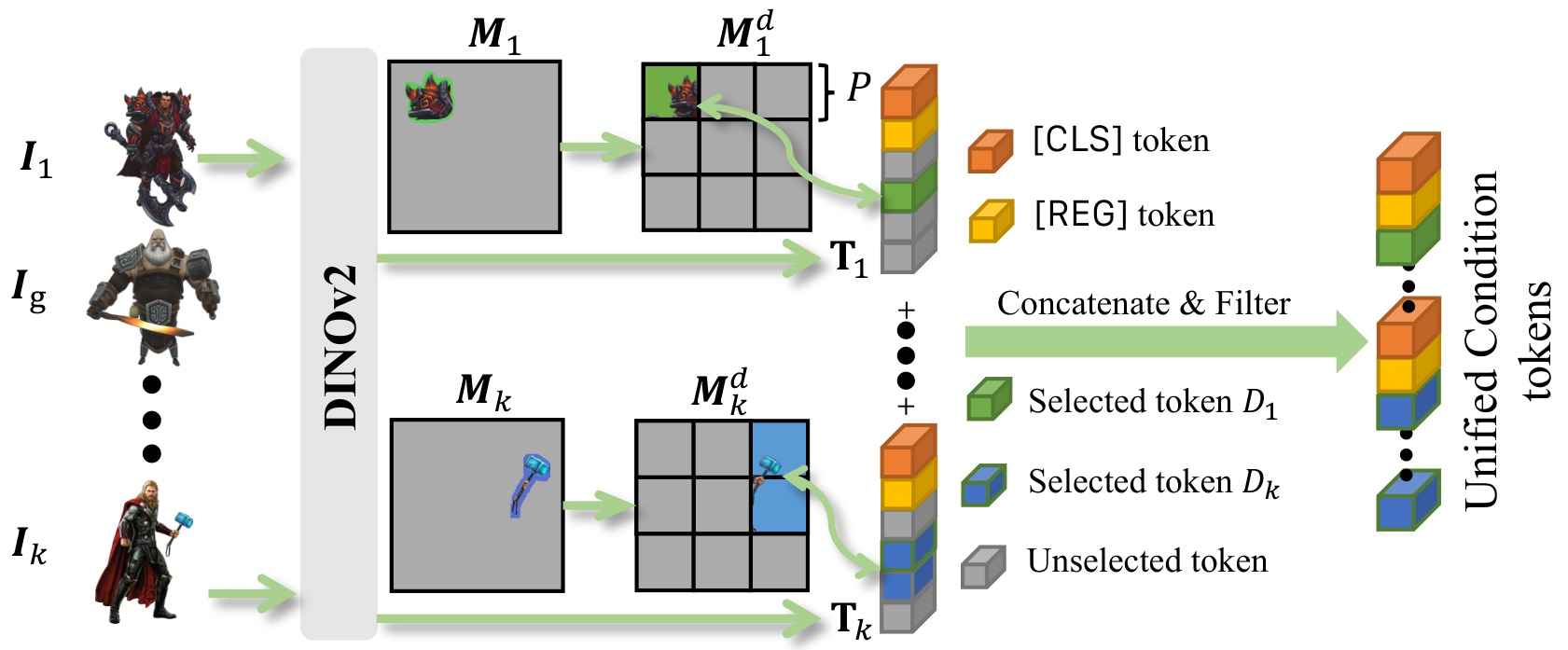}
    \vspace{-0.3cm}
    \caption{
       The processing pipeline of MCFM to fuse multiple regions.
    }
   \label{fig:MCFM}
    \vspace{-0.5cm}
\end{figure}


Given the input consists of a global condition image $I_{\text{g}}$ and a set of $K$ local condition images with user-specified 2D binary masks:
\[
\mathcal{P} = \{(\boldsymbol{I}_k, \boldsymbol{M}_k)\}_{k=1}^{K}, \quad \boldsymbol{M}_k \in \{0,1\}^{H \times W},
\]
where $k$ is the index of the local condition images, $\boldsymbol{M}_k$ is the selected regions of the $k$th image $\boldsymbol{I}_k$.

Our goal in this section is to design a fusion module that merges multiple control features from 2D regions into unified condition token to guide subsequent 3D asset generation.
Since TRELLIS relies on DINOv2~\cite{dinov2, dinoReg} for feature extraction, we adopt DINOv2 as our token encoder to ensure that the fused condition tokens lie in the same latent space. This design allows us to fully leverage the pretrained capacity of TRELLIS in the subsequent generation process.

However, we observe that directly cropping each 2D region in image space and feeding their independently encoded features into the generation pipeline often results in failure when guiding the synthesis of a coherent 3D asset.
We attribute this to the fact that cropped regions lose two critical informations.
First, removing the full image context loses the relative positional encoding provided by DINOv2.
Second, without the complete image, these local condition regions can no longer interact globally through self-attention. Both of the above play a pivotal role in enabling the model to reason about the spatial relationships within the condition images.
As a result, the encoded tokens from cropped images lack consistent global context.

To address this, we propose a Multi-Condition Fusion Module (MCFM) that integrates multiple regions without disrupting their spatial alignment or global coherence. 

As shown in~\refFig{fig:MCFM}, each local condition image $\boldsymbol{I}_k$ is directly fed into the DINOv2 model without cropping out the unselected regions.
During the encoding process, image $\boldsymbol{I}_k$ is divided into non-overlapping $P \times P$ patches, and each patch is encoded as a token. Moreover, a \texttt{[CLS]} token and several \texttt{[REG]} tokens will be prepended to the patch tokens sequence.
\texttt{[CLS]} token serves as a global summary token for classification, while \texttt{[REG]} tokens are designed to capture additional global context~\cite{dinov2, dinoReg}.
After encoding, each condition image $\boldsymbol{I}_k$ produces the following token sequence:
\vspace{-0.5em}
\[
\mathbf{T}_k:( \texttt{CLS}, \texttt{REG}_1,\ldots ,\texttt{REG}_r, \text{token}_1, \ldots, \text{token}_{\tilde{H}\tilde{W}})_k,
\]
where $\tilde{H} = H / P$ and $\tilde{W} = W / P$, H and W are the resolution of the condition image.

As each token corresponds to one original patch in the image space (excluding special tokens). We can simply select the corresponding condition feature tokens using the 2D region mask $\boldsymbol{M}_k$. We divide $\boldsymbol{M}_k$ into mask $\boldsymbol{M}^d_k \in \{0,1\}^{\tilde{H} \times \tilde{W}}$ that correspond to the number of patches. 
Then the selected indexes of condition tokens $\boldsymbol{D}_k$ are given by:
\[
\boldsymbol{D}_k = \left\{ i \mid \boldsymbol{M}^d_k[i // \tilde{W},\ i \bmod \tilde{W}] = 1 \right\},
\]
where $i$ is the indices of flattened patch sequence.

Then, for each $\mathbf{T}_k$, we retain the \texttt{[CLS]} token, the \texttt{[REG]} tokens, and the patch tokens indexed by $\boldsymbol{D}_k$.
These retained tokens from all condition images are then concatenated to form the unified condition tokens.
Importantly, this concatenation does not violate the structure of the downstream generation model, as these condition tokens are injected via cross-attention rather than as sequential inputs to a transformer encoder. Consequently, the number of tokens along the first dimension remains scalable.

\subsection{3D Semantic-Aware Alignment }\label{method:3D_Align}

\begin{figure}[!htp] 
    \centering
    \includegraphics[width=\linewidth]{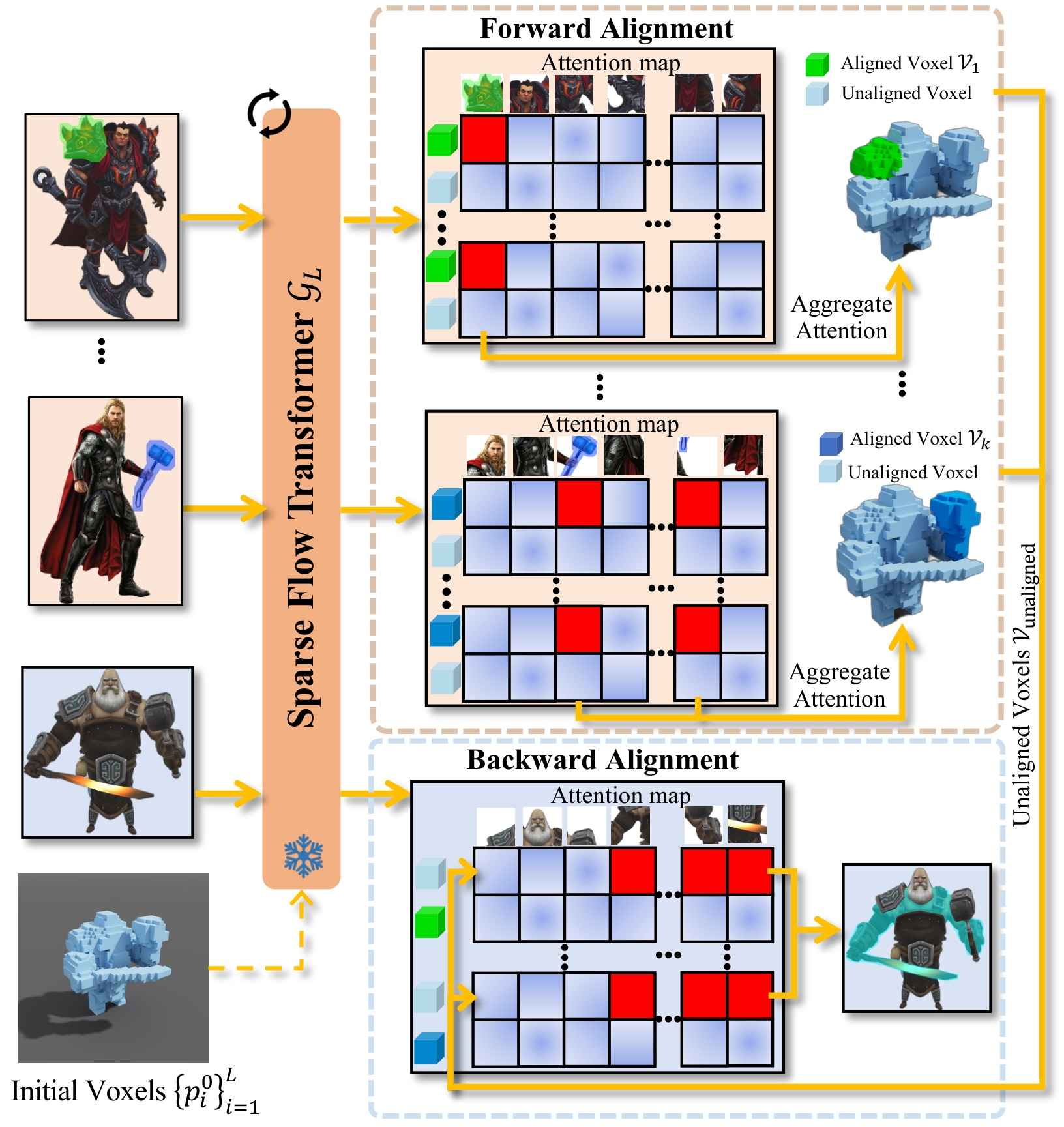}
     
    \caption{
        The processing pipeline of 3D Semantic-Aware Alignment to to identify the 2D-to-3D correspondences.
    }
   \label{fig:3D_aware}
   \vspace{-2.5em}
\end{figure}

Equipped with control feature tokens from multiple conditional images, we need to further establish correspondence between the tokens and the 3D structure, which enables precise local control over the 3D assets. To this end, we propose a 3D Semantic-Aware Alignment in this section as shown in~\refFig{fig:3D_aware} , which includes a bidirectional token-to-voxel and voxel-to-token matching pass.

\paragraph{Forward Alignment: Image-Guided Voxel Selection.}

We observe that the pretrained transformer-based flow model $\boldsymbol{\mathcal{G}}_{\mathrm{L}}$, originally designed to generate latent for active voxels conditioned on image tokens, inherently learns a Semantic-Aware alignment between 2D image content and 3D voxel locations.
Leveraging this property, we use $\boldsymbol{\mathcal{G}}_{\mathrm{L}}$ to establish correspondences between user-selected 2D regions and their associated 3D regions.

For each local condition image, we feed the full extracted tokens into $\boldsymbol{\mathcal{G}}_{\mathrm{L}}$ to generate latents for initial voxels $\{\boldsymbol{p}_i^0\}_{i=1}^{L}$.
During this generation, the image tokens are injected through cross-attention layers as keys and values, while the latent attached on the voxels $\{\boldsymbol{p}_i^0\}_{i=1}^{L}$ act as queries.
This produces a cross-attention map between voxels and image tokens.
Note that this cross-attention map is normalized via softmax
along the image tokens axis, so the score attached to a voxel
quantifies how strongly that voxel aligns with the selected token.
We aggregate attention scores over the selected tokens indexed by mask $\boldsymbol{D}_k$ and apply a threshold to identify the set of voxels that are semantically aligned with the selected region $\boldsymbol{M}_k$ on the condition image.
This process is independently repeated for each condition image, resulting in a set of aligned voxels
\[
\{\boldsymbol{p}_i|i\in\mathcal{V}_k\}_{k=1}^K,
\]
where $\mathcal{V}_k$ denotes the index of aligned voxels for $\boldsymbol{I}_k$. After getting the aligned voxels, we apply a simple voting strategy to fill small holes and remove isolated outliers, which will be illustrated in the supplementary material.

\paragraph{Reverse Alignment: Voxel-Guided Image Patch Selection.}
To identify which tokens from the global condition image correspond to the remaining unaligned voxels, we apply the same mechanism as above to produce a cross-attention map between the global condition image tokens and voxels.
We then identify the unaligned voxels:
\vspace{-0.5em}
\[
\{\boldsymbol{p}_i|i\in\mathcal{V}_{\text{unaligned}}\},\quad 
\mathcal{V}_{\text{unaligned}} = \{1, \ldots,L \} \setminus \bigcup_{k=1}^K \mathcal{V}_k,
\]
where $\mathcal{V}_{\text{unaligned}}$ is the index of the unaligned voxels.
By aggregating attention scores over the unaligned voxels, we identify the tokens that correspond to the unaligned voxels.
During the reverse alignment, the cross-attention matrix is normalized via softmax along the voxel axis. 
Summing those weights over the unaligned voxels $\mathcal{V}_{\text{unaligned}}$ gives a score that measures how strongly that token is aligned with the unaligned voxels.
We denote the index set of these selected tokens of the global condition image as $\boldsymbol{D}_{\text{g}}$, following the same selection method used in the forward alignment. Note that the global condition tokens $\boldsymbol{D}_{\text{g}}$ are concatenated with the local condition tokens $\boldsymbol{D}_k$ to form the final unified condition token.

\subsection{Local Attention Enhancement}\label{Sec:localEnhance}

\begin{figure}[!htp]
    \centering
    \includegraphics[width=\linewidth]{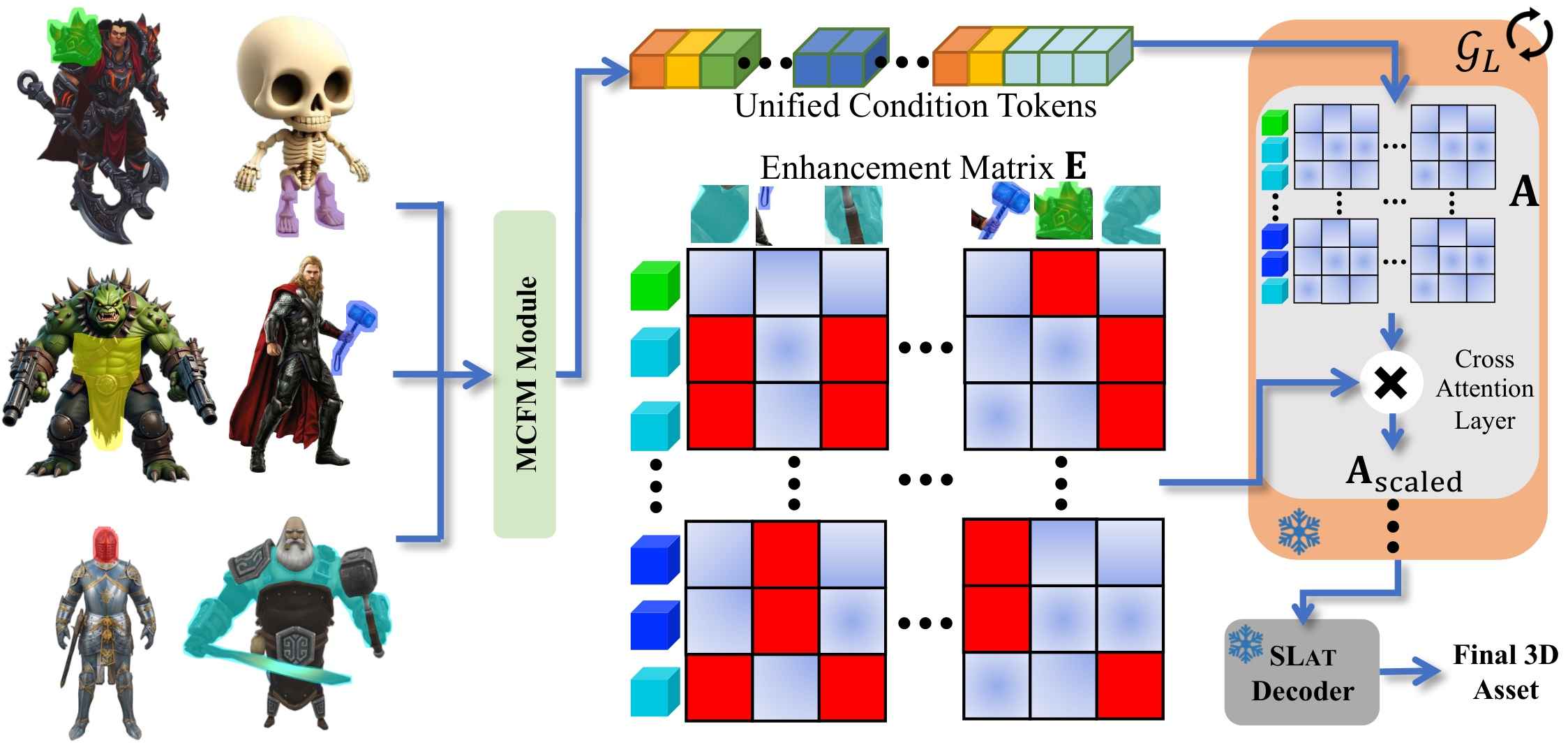}
    \vspace{-1em}
    \caption{
        The pipeline of local attention enhancement module.
    }
   \label{fig:localEnhance}
   \vspace{-1.8em}
\end{figure}




To achieve region-specific 3D generation, we leverage the condition tokens obtained in ~\refSec{sec:MCFM} and~\refSec{method:3D_Align}, along with the 2D-3D correspondence, to perform inference. However, directly applying these tokens during generation can lead to unintended blending, where features from different condition images influence the same 3D regions.
While in some cases, blending features across multiple images is actually desirable, especially when users aim to create unified styles or shared visual features.
Thus, we introduce a local attention enhancement strategy as shown in~\refFig{fig:localEnhance} that adjusts the fusion strength for local control features.

Specifically, the unified condition tokens are injected as keys and values into the cross-attention layers of $\boldsymbol{\mathcal{G}}_{\mathrm{L}}$, while the latent features attached to all voxels $\{\boldsymbol{p}_i^0\}_{i=1}^{L}$ serve as queries.  
This results in a cross-attention map $\mathbf{A}$, where each row corresponds to a voxel and each column to an image token.


To control the spatial influence of each condition image, we calculate a attention enhancement matrix 
$\mathbf{E}$ applied to the cross-attention map $\mathbf{A}$ before the softmax operation.
For each local image $\boldsymbol{I}_k$, we assign an enhance strength $\lambda_k$ to control its contribution.
For each local image $\boldsymbol{I}_k$, and for every token index $j \in \boldsymbol{D}_k$ and voxel index $i \in \mathcal{V}_k$, we set:
$
\mathbf{E}[i, j] = \lambda_k.
$

The same operation is applied to the global image $\boldsymbol{I}_{\text{g}}$ using $\boldsymbol{D}_{\text{g}}$ and $\mathcal{V}_{\text{unaligned}}$.
All other entries in $\mathbf{E}$ are set to 1. 
The enhanced attention logits are computed as:
$\mathbf{A}_{\text{scaled}} = \mathbf{A} \odot \mathbf{E}.$

This mechanism modulates how strongly each token influences its aligned 3D region.
Lower $\lambda$ values promote smoother blending across regions, while higher values enforce more localized, disentangled control.
By default, those $\lambda_k$  are determined by the proportion of the selected region in the entire condition image. 
This design ensures that more precisely selected 2D regions (i.e., those with fewer selected tokens relative to the full image) exert a stronger and more localized influence on the corresponding 3D voxels.
\vspace{-0.8em}
\section{experiment}
\begin{figure*}
  \centering
  \includegraphics[width=\textwidth]{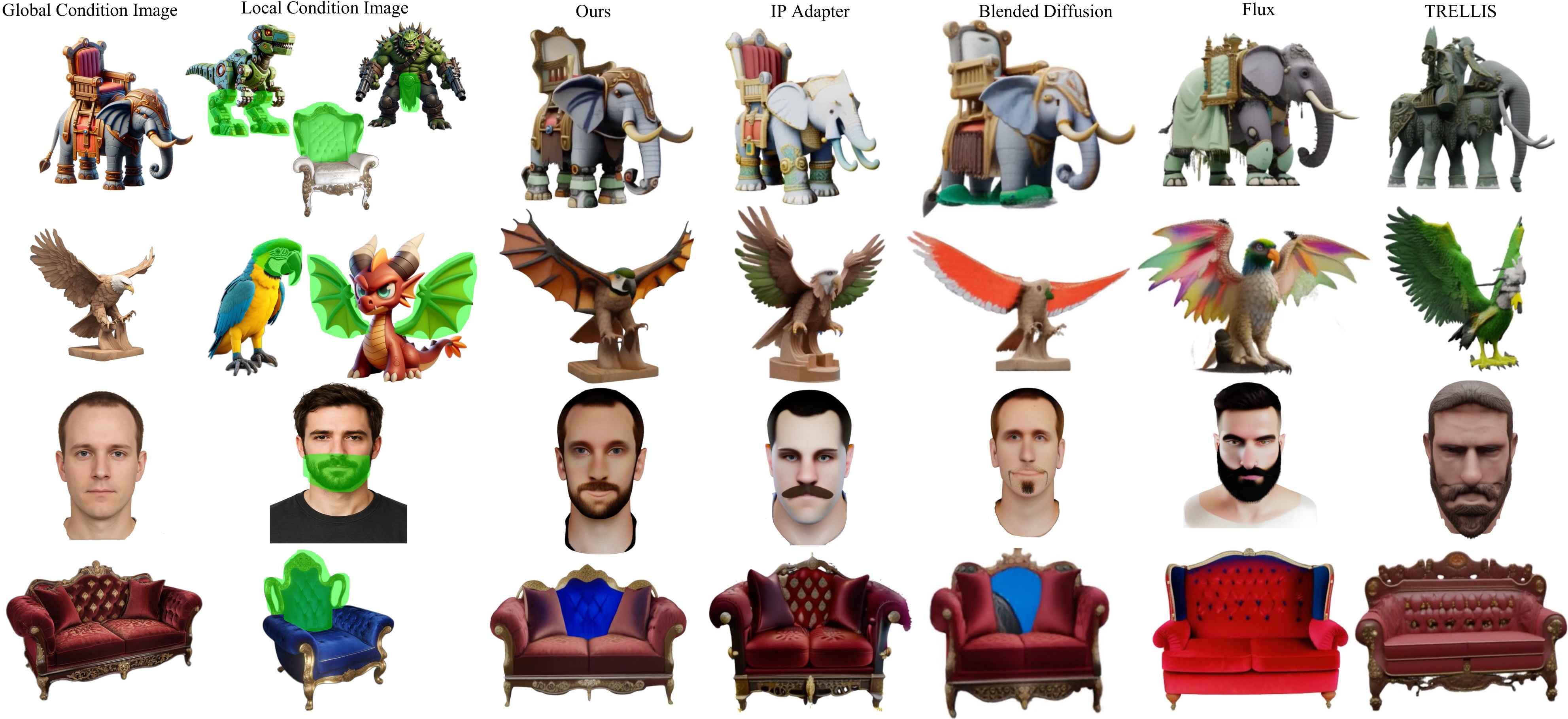}
  \vspace{-1.5em}
  \caption{
  Generation quality comparison with previous methods.
  }
  \label{fig:mainCompare}
    
\end{figure*}

\subsection{Experimental Setting}

\paragraph{Evaluation Metrics}

We evaluate the alignment between the generated 3D assets and the provided conditions using \revise{CLIP-based metrics}~\cite{clip}. 
For \revise{global CLIP similarity}, we employ GPT-4o~\cite{gpt4o} to automatically generate a caption that summarizes the intended fusion of the global and local condition images. 
The generated caption is then used as the textual input for computing CLIP similarity, with details of the captioning procedure provided in the supplementary material. 
\revise{Beyond global similarity, we report region-specific CLIP similarity, which directly measures part-level controllability by comparing masked renderings of the generated 3D asset with the corresponding masked local condition regions. 
This provides a finer-grained evaluation of whether each local input is faithfully preserved.}

In addition, following A3D~\cite{A3D}, we further adopt GPTEval3D~\cite{gptEval} to evaluate the perceptual quality and the controllability of the generated 3D assets. GPT-4o serves as a fair user proxy to evaluate the performance of various methods.
Specifically, we provide GPT-4o with two results generated by different methods along with a textual question for evaluation, and then ask it to select which result performs better under each evaluation criterion.
The evaluation includes five criteria, which are region consistency, visual seamlessness, edit controllability, detail preservation, and overall preference.
The detailed investigation process can be found in the supplementary material.

Finally, we calculate ImageReward~\cite{imageReward, coin3d} scores, which are computed by a learned reward model trained to align with human preferences, providing a scalar estimate of the perceptual quality of each result.
\vspace{-0.4em}

\paragraph{Baselines}

\revise{Since no existing method directly supports multi-image, region-level fusion for 3D generation, constructing fair baselines is challenging. 
To approximate our setting, we employ GPT-4o to generate a structured textual description that summarizes the intended fusion, which is then used to guide baseline methods toward the same objective.}
We compare \name{} with four different methods. \revise{Additional implementation details and comparisons with further baselines are provided in the supplementary material.}
\textbf{(1)} IP-Adapter~\cite{ipAdapter} is a lightweight adapter module that enables diffusion models to be conditioned on both textual prompts and image prompts simultaneously.
We provide the textual description and the global conditioning image as input conditions to IP-Adapter to synthesize a fused image.
This image is then passed to TRELLIS to obtain the final 3D asset.
This setup reflects a multi-condition fusion strategy similar in spirit to our method.
\textbf{(2)} \revise{Blended Diffusion~\cite{blendedDiffusion} performs text-guided inpainting on the global image using user-defined masks, and the fused result is passed to TRELLIS.}
\textbf{(3)} FLUX~\cite{flux} is a powerful text-to-image model that supports high-fidelity generation guided by prompts. We use the textual description as input to FLUX, and then use TRELLIS to generate the corresponding 3D asset.
\textbf{(4)} TRELLIS~\cite{trellis} provides a text-to-3D generation pass. We use the textual description to directly generate the 3D asset.

\begin{table}[!htp]
\centering
\caption{
Quantitative comparison across different methods.
}
\vspace{-1em}
\resizebox{\linewidth}{!}{
    \begin{tabular}{lccccc}
    \toprule
    Metric & IP-Adapter & \revise{Blended Diffusion} & FLUX & TRELLIS & Ours \\
    \midrule
    \revise{Region-specifc CLIP Similarity} $\uparrow$ & 0.223 & 0.240 & 0.224 & 0.237 & \textbf{0.243} \\
    Global CLIP Similarity $\uparrow$ & 0.201 & 0.236 & 0.238 & 0.235 & \textbf{0.241} \\
    ImageReward $\uparrow$ & -1.21 & -0.79 & -0.54 & -0.63 & \textbf{-0.51}\\
    \bottomrule
    \end{tabular}
}
\vspace{-1em}
\label{tab:CLIPcompare}
\end{table}
\begin{table}
\centering
\caption{
Quantitative comparison for the controllable generation. Values greater than 50 indicate that our method achieves better results.
}
\vspace{-1em}
\setlength{\tabcolsep}{2.6mm}
\small
\begin{tabular}{lccccc}
\toprule
Method & \multicolumn{5}{c}{
\begin{tabular}{@{}c@{}}
GPTEvals3D $\uparrow$, \% of comparisons \\ where our method is preferred
\end{tabular}}
\\ 
\hline
{} & \textbf{R. C.} & \textbf{V. S.} & \textbf{E. C.} & \textbf{D. P.} & \textbf{O. P.}  \\ 
vs. IP-Adapter & 84 & 78 & 76 & 92 & 76 \\
\revise{vs. Blended Diffusion} & 63 & 73 & 59 & 68 & 60 \\
vs. FLUX & 55 & 56 & 63 & 52 & 55\\
vs. TRELLIS & 70 & 67 & 70 & 74 & 68 \\
\bottomrule
\end{tabular}
\label{tab:GPTcompare}
\vspace{-1em}
\end{table}

\subsection{Comparison Results}
\paragraph{Qualitative Evaluation}

As shown in \refFig{fig:mainCompare}, we visualize the generation result of different methods.
Compared to all baselines, our method better combines visual elements from the selected condition regions with the overall feature from global image, maintaining both structure and local appearance.
IP-Adapter shows weak localized adjustment capabilities. In the Hulk and sofa examples, it visually leaves the sofa and Hulk almost unchanged, failing to apply any noticeable edits based on the provided text.
Ctrl-X does not provide region-specific control, and all the local control is broadly derived from a single appearance image. Consequently, while the geometry is approximately correct, the appearance control lacks precision.
FLUX and TRELLIS only conditioned on textual input, which often lacks sufficient detail for accurate 3D generation. For example, the generated Hulk does not match the intended structure from the global image.

As shown in \refFig{fig:mainCompare}, we visualize the generation result of different methods.
Compared to all baselines, our method better combines visual elements from the selected condition regions with the overall feature from global image, maintaining both structure and local appearance.
IP-Adapter shows weak localized adjustment capabilities. In the sofa examples, it visually leaves the sofa almost unchanged, failing to apply any noticeable edits based on the provided text.
\revise{Blended Diffusion relies on manual-drawn masks, so the edited location is correct. However, the edit is driven purely by text rather than the local image, leading to imprecise attribute transfer. In the face example it can add a beard, but its shape and boundaries are inaccurate.}
FLUX and TRELLIS only conditioned on textual input, which often lacks sufficient detail for accurate 3D generation. For example, the generated human face does not match the intended structure from the global image.

\begin{table}[!htp]
\vspace{-1.2em}
\centering
\caption{
Quantitative results for ablation study.
}
\vspace{-1em}
\resizebox{\linewidth}{!}{
    \begin{tabular}{lcccc}
    \toprule
    Metric & w/o L.E. & w/o MCFM & w/o R.A. & Ours \\
    \midrule
    \revise{Region-specific CLIP Similarity} $\uparrow$ & 0.220 & 0.156 & 0.199 & \textbf{0.227}  \\
    Global CLIP Similarity $\uparrow$ & 0.218 & 0.176 & 0.179 & \textbf{0.223}  \\
    ImageReward $\uparrow$ & -0.85 & -1.01 & -1.17 & \textbf{-0.79}\\
    \bottomrule
    \end{tabular}
}
\label{tab:CLIPAbla}
\vspace{-1.5em}
\end{table}

\paragraph{Quantitative Evaluation}

We evaluate our method against four baselines using GPTEval3D, CLIP similarity and ImageReward.
As shown in \refTab{tab:GPTcompare}, our method achieves the highest preference across all five GPTEval3D metrics: Region Consistency, Visual Seamlessness, Edit Controllability, Detail Preservation, and Overall Preference, demonstrating superior controllability and visual quality.
In addition, \refTab{tab:CLIPcompare} shows that we also outperform all baselines in both CLIP Similarity and ImageReward scores.

\subsection{Ablation Study}

We perform ablation studies to assess the effectiveness of our components.
The quantitative results, as shown in \refTab{tab:CLIPAbla}, demonstrate that removing any single component consistently leads to lower CLIP Similarity and ImageReward scores, highlighting the contribution of each module to the overall generation quality.

\begin{figure*}
  \centering
  \includegraphics[width=\linewidth]{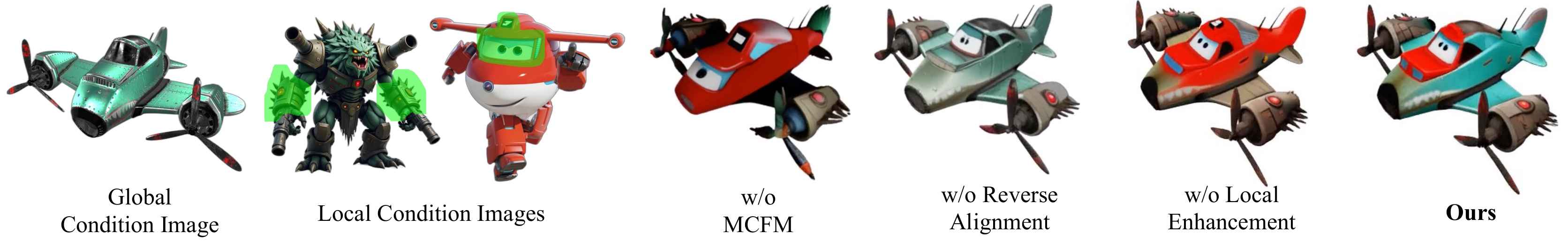}
  \caption{
  We illustrate how each proposed module contributes to final generation result.
  }
  \label{fig:ablaAll}
\end{figure*}

In addition, we show the qualitative results in \refFig{fig:ablaAll} to analyze how each module contributes to controllable 3D generation.
When the MCFM is removed, the model can not infer spatial relationships (e.g., the two "eyes" of the aircraft belong to the front window in \refFig{fig:ablaAll}).
Without 3D Semantic-Aware Alignment, the 3D regions are influenced by both the global and local images, leading to chaotic target 3D fusion results(e.g., in the 4th column in \refFig{fig:ablaAll}, the front window retains the 'eyes' from the global image but loses the red outer shell specified by the local image). Disabling Local Enhancement Generation causes features from different sources to bleed across region boundaries. We provide more discussions and results about the ablation study in the supplementary material.

\section{Conclusion}
In this work, we propose \name, a novel framework for controllable 3D asset generation guided by multiple condition images with user-selected regions.
In order to achieve multi-condition fusion control, we propose Multi-Condition Fusion Module (MCFM) enables the integration of visual cues from multiple condition images into a unified condition tokens representation. Then, we propose the 3D Semantic-Aware Alignment Module, which extracts attention information from the pretrained TRELLIS to establish 2D-3D correlations, enabling the condition token to precisely control the target 3D structure. Finally, we propose Local Attention Enhancement, which constructs a local attention matrix to resolve conflicts from multiple control targets on the same 3D structure, ensuring distinct local features and smooth component integration.
Extensive experiments and ablation studies demonstrate the effectiveness and superiority of our approach over existing methods in producing highly controllable and visually convincing 3D assets.


\begin{acks}
This work was partially supported by the National Key R\&D Program of China (No. 2024YDLN0011), NSFC (No. 62572436), NSFC Corporate Joint Key Program (U22B2034), and the Key R\&D Program of Zhejiang Province (No. 2023C01039).
\end{acks}

\bibliographystyle{ACM-Reference-Format}
\bibliography{main}

\begin{figure*}
    \centering
    \includegraphics[width=\textwidth]{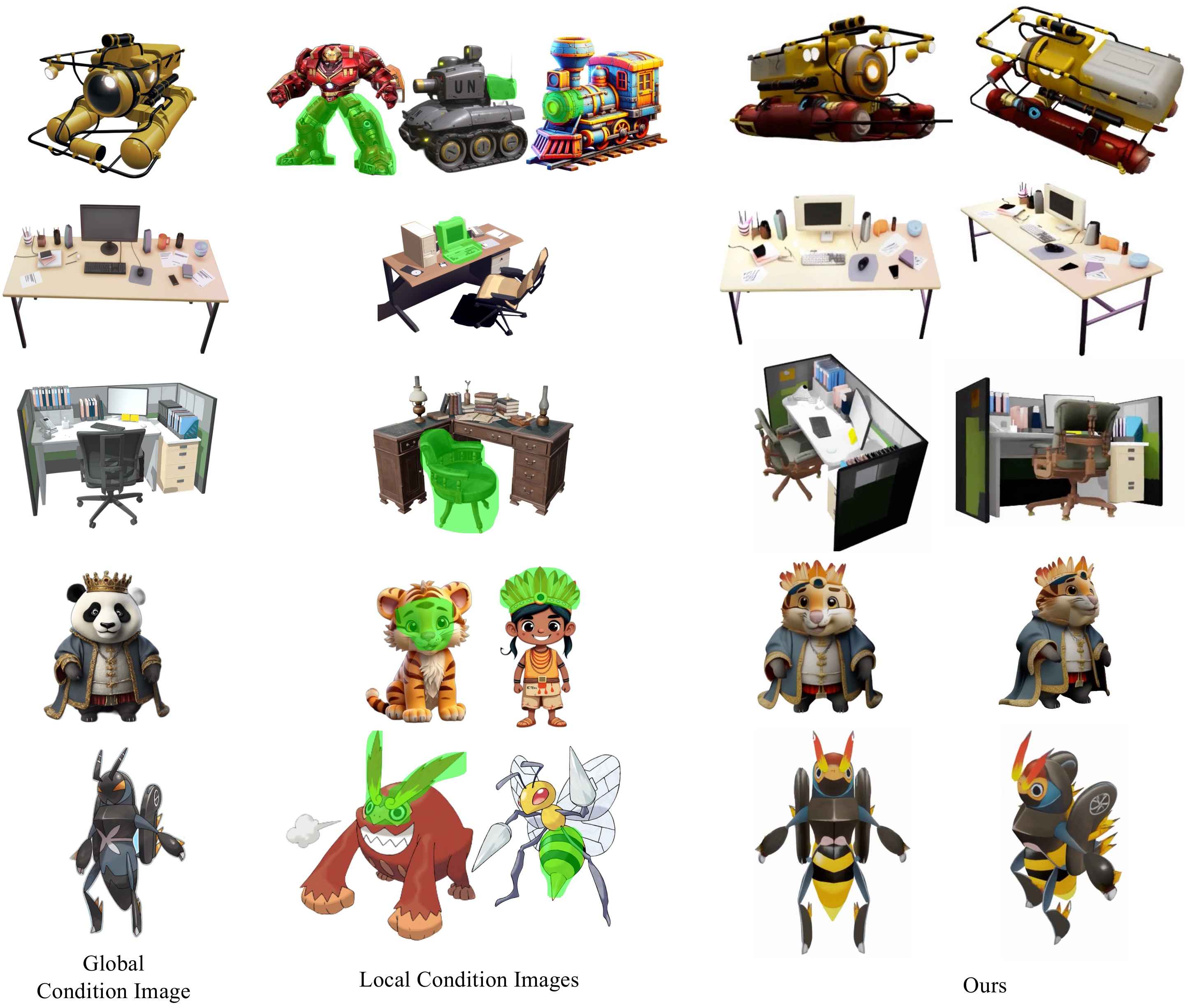}
    \caption{
        More results of controllable 3D asset generation.
    }
   \label{fig:result0}
\end{figure*}

\begin{figure*}
    \centering
    \includegraphics[width=\textwidth]{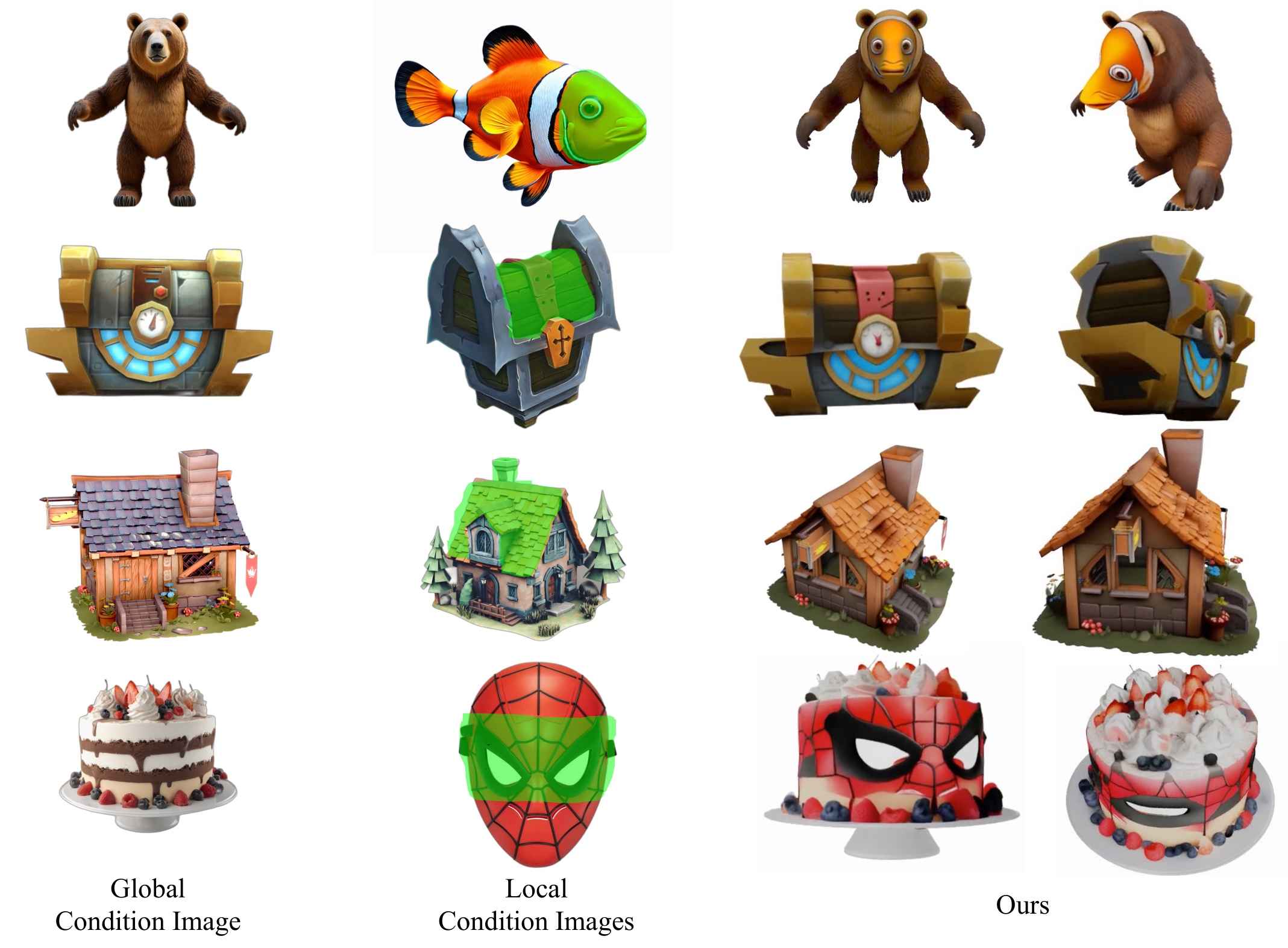}
    \caption{
        More results of controllable 3D asset generation.
    }
   \label{fig:result0}
\end{figure*}

\begin{figure*}
    \centering
    \includegraphics[width=\textwidth]{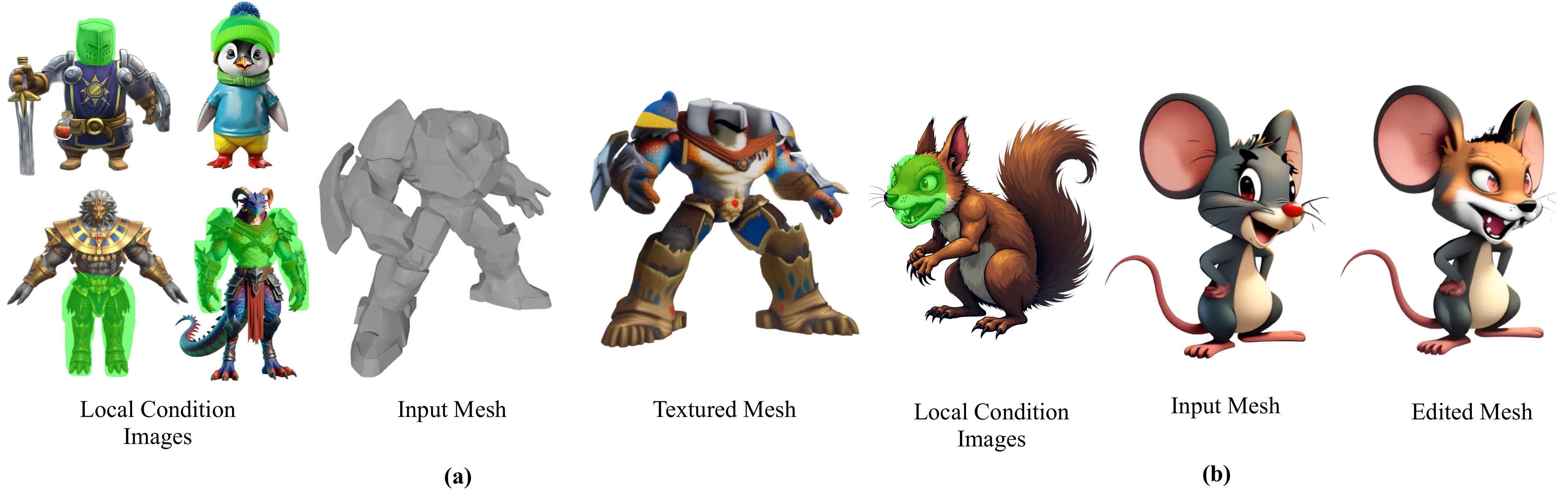}
    \caption{
        Applications: \textbf{(a):} Given an untextured mesh, our method generates a coherent textured result by fusing multiple local condition images, preserving diverse features.
        \textbf{(b):} Our approach enables localized mesh editing by transferring visual features from selected 2D regions onto the target mesh, producing semantically meaningful and stylistically aligned modifications.
    }
   \label{fig:result0}
\end{figure*}
\clearpage
\appendix

\twocolumn[
\begin{center}
    {\huge \bf Fuse3D: Generating 3D Assets Controlled by Multi-Image Fusion}\\
    {\LARGE \emph{(Supplementary Material)}}
\end{center}
\vspace{0.1cm}
]

\section{More Method Details}

\subsection{Selected Voxel Refinement}\label{app:voxelRefine}

\begin{figure}[!htp]
  \centering
  \includegraphics[width=\linewidth]{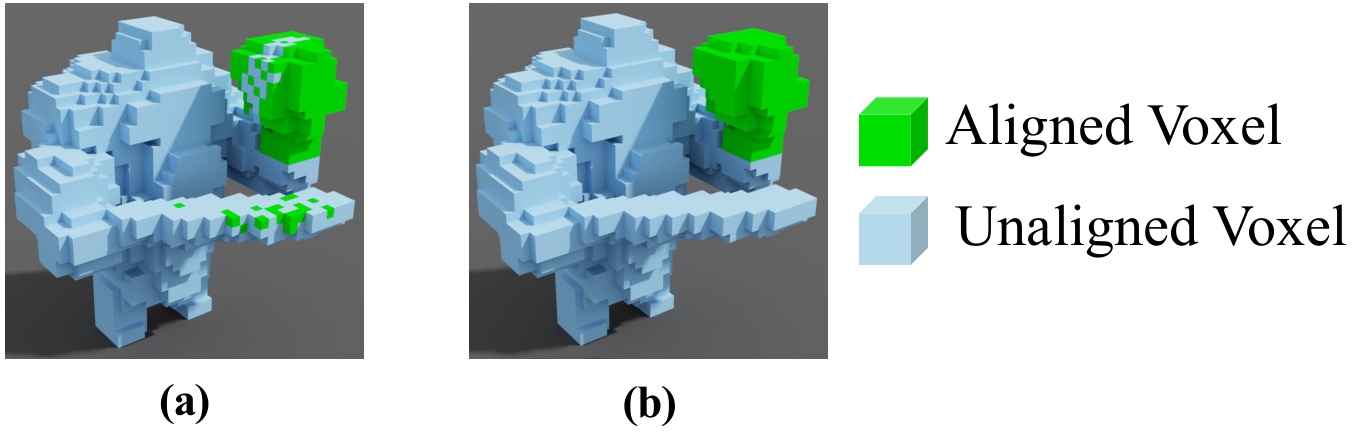}
  \caption{
  \textbf{(a):} Raw voxel mask obtained by thresholding the attention scores.
  \textbf{(b):} Voxel mask after the refinement.
  }
  \label{fig:Supp_knn}
\end{figure}

In Section 3.3 of the main text, we present the method of constructing correspondences between condition tokens and 3D voxels using 3D Semantic-Aware Alignment.
Though the raw voxel mask obtained by thresholding the aggregated attention scores is largely correct, due to its fully automatic nature, the matching results may contain pinholes inside the selected area and isolated voxels at the boundary as shown in~\refFig{fig:Supp_knn}.a.
We repair it with a simple majority vote rule in the voxel lattice.
For every voxel, we collect the indices of its k nearest neighbors ( k = 16 in practice ) and count how many of them are marked as selected.
A previously unmarked voxel is filled if at least 60 \% of its neighbors are selected;
a previously marked voxel is cleared if fewer than 40 \% of its neighbors are selected.
The operation is applied once and yields a topologically coherent region that exactly follows the user-intended area while eliminating outliers.
A visual comparison of the raw and refined voxel masks is provided in ~\refFig{fig:Supp_knn}, where \textbf{(a)} shows the initial noisy alignment with scattered outliers, and \textbf{(b)} demonstrates how our refinement produces a cleaner voxel mask.

\begin{figure*}[htbp]
  \centering
  \includegraphics[width=\linewidth]{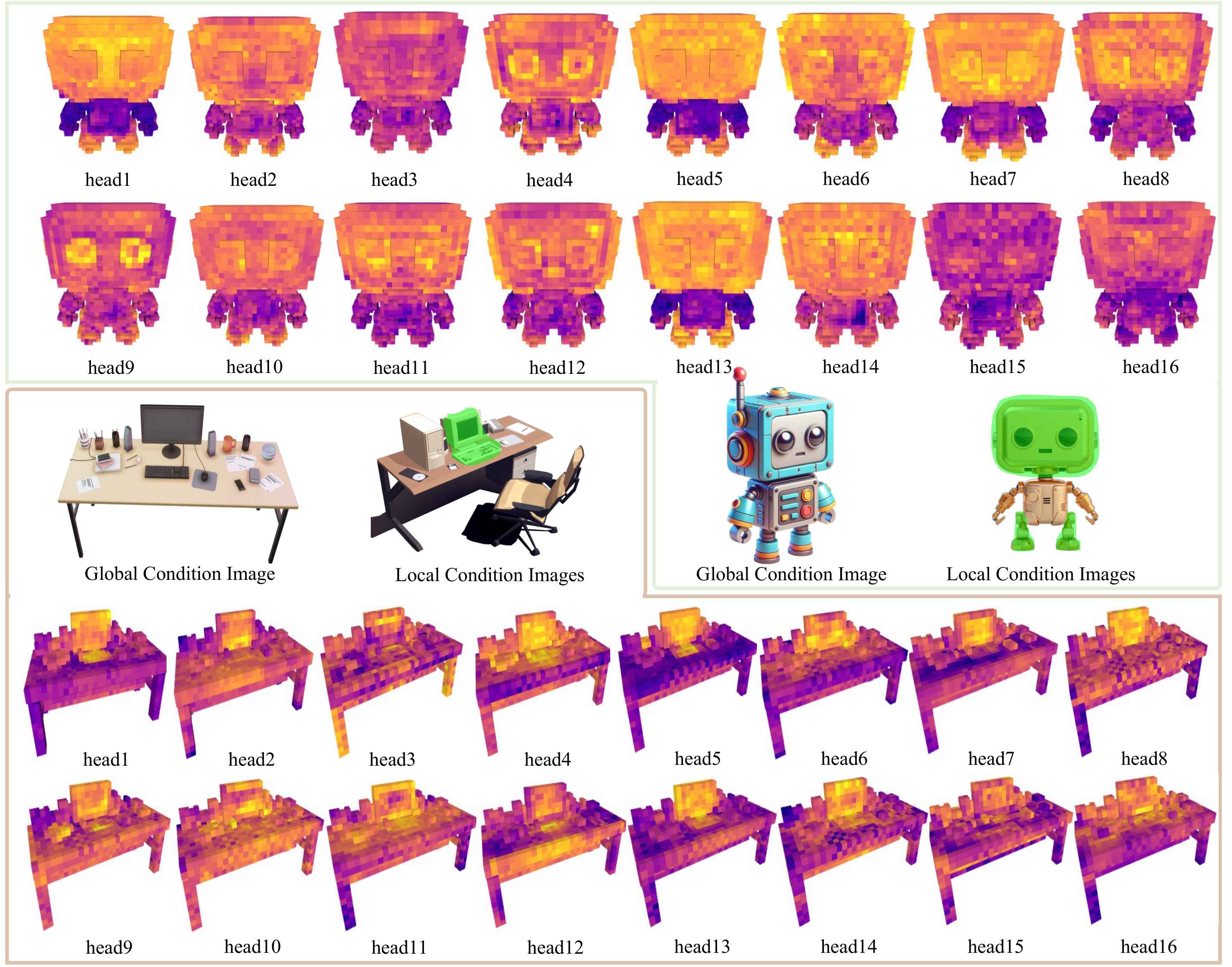}
  \caption{
  Visualization of attention scores from different cross-attention heads during the forward process of 
    $\mathcal{G}_\mathrm{L}$  }
  \label{fig:Supp_heads}
\end{figure*}

\subsection{Attention Scores of Different Heads}

During the 3D-aware alignment process, we rely on the cross-attention maps produced by the image-conditioned flow model $\boldsymbol{\mathcal{G}}_{\mathrm{L}}$ to establish correspondences between selected 2D regions and 3D voxels.
However, we observe that the attention scores vary significantly across different attention heads within the same cross-attention layer, especially in their ability to correctly localize the intended semantic region in 3D space.
This functional divergence across heads has also been observed in prior works~\cite{attentionHeads}.

To better understand this phenomenon, we visualize the attention scores from all 16 heads for a specific input case in \refFig{fig:Supp_heads}.
Each heatmap illustrates the spatial distribution of attention across the voxel grid for one head.
As shown, certain heads such as head 1, head 5, and head 13 produce much sharper and semantically aligned attention regions, while other heads yield more diffuse or noisy responses.

Based on this observation, instead of aggregating attention scores uniformly across all heads, we select the subset of heads 1, 5 and 13 that consistently produce focused and semantically meaningful activations.
We then sum their logits to compute the final alignment score used for voxel selection.


\subsection{\revise{Implementation Details}}
\revise{
In Fuse3D, we adopt TRELLIS~\cite{trellis}, specifically trellis-image-large version, as it can generate impressive 3D assets from image and text prompts. We did not perform additional fine-tuning on the TRELLIS model. Instead, we utilized certain components from it along with our proposed modules to construct the complete Fuse3D pipeline. It is worth noting that our method is not limited to TRELLIS. With minor modifications, it can be adapted to other 3D native generation models, such as CLAY~\cite{clay}. We will release all Fuse3D code to support community development. 
}

\revise{
All experiments in this paper were conducted on a single NVIDIA A6000 GPU. Under typical circumstances, fusing three condition images to generate the corresponding 3D assets takes less than 20 seconds. There is considerable potential to optimize the timing further, for example, by adopting faster cross-attention computation techniques or employing more powerful GPUs. Notably, adding more condition images does not significantly increase inference time, as Fuse3D can process each condition image in parallel.
}

\section{Experiment Detials}

\subsection{\revise{Hyperparameters}}

\revise{
We set the attention score threshold in Sec.~3.3 to \textbf{0.55}. The patch size for feature extraction follows TRELLIS and is fixed to \textbf{14}.}

\begin{figure*}
  \centering
  \includegraphics[width=\textwidth]{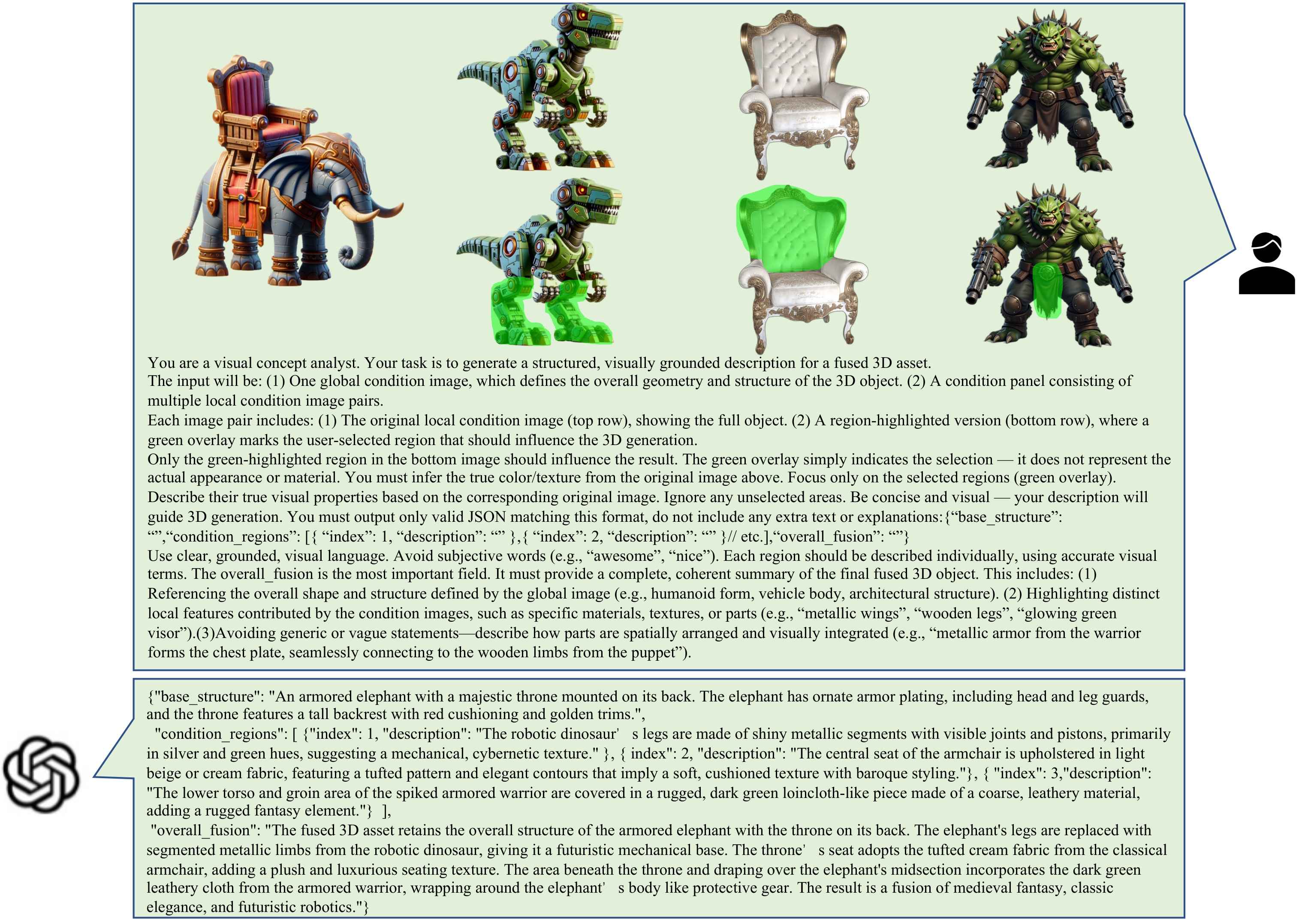}
  \caption{
  An example of the caption process.
  }
  \label{fig:Supp_caption}
\end{figure*}
\subsection{Caption Process}\label{Supp_caption}

Because our method is the first to enable the fusion of multiple 2D condition images for controllable 3D generation, existing captioning frameworks are not directly applicable for evaluation. To conduct fair comparisons with other methods, we design a custom prompt that explicitly describes the input setting of our task. The obtained captions will be used as input for the baseline methods and for the computation of metrics.

Unlike prior captioning approaches~\cite{VFC}, which typically describe a single 3D asset or a single image, our inputs consist of multiple condition images with user-selected regions. These selected regions jointly contribute to the generation process and must be properly reflected in the description.

To address this, we design a structured prompt tailored for GPT-4o, guiding it to generate a detailed and grounded textual description of the expected fused 3D asset based on the base image and the selected 2D regions from each condition image. This prompt ensures consistency across methods and enables alignment-based metrics such as CLIP similarity and ImageReward.
An illustration of our captioning pipeline is provided in \refFig{fig:Supp_caption}.

\subsection{Metrics}

\paragraph{CLIP}

\revise{We adopt the ViT-L/14 variant of CLIP~\cite{clip} and report two types of similarity scores. 
First, global CLIP similarity measures the alignment between the RGB rendering of the generated 3D asset and the fused textual description obtained from GPT-4o. 
Second, region-specific CLIP similarity directly evaluates part-level controllability: for each local condition, we render the corresponding masked region of the 3D asset and compute CLIP similarity against the corresponding text description.}

\paragraph{ImageReward}

We adopt ImageReward-v1.0~\cite{imageReward} to compute the perceptual reward between the RGB rendering of the 3D asset and its corresponding textual description.

\begin{figure*}[htbp]
  \centering
  \includegraphics[width=\linewidth]{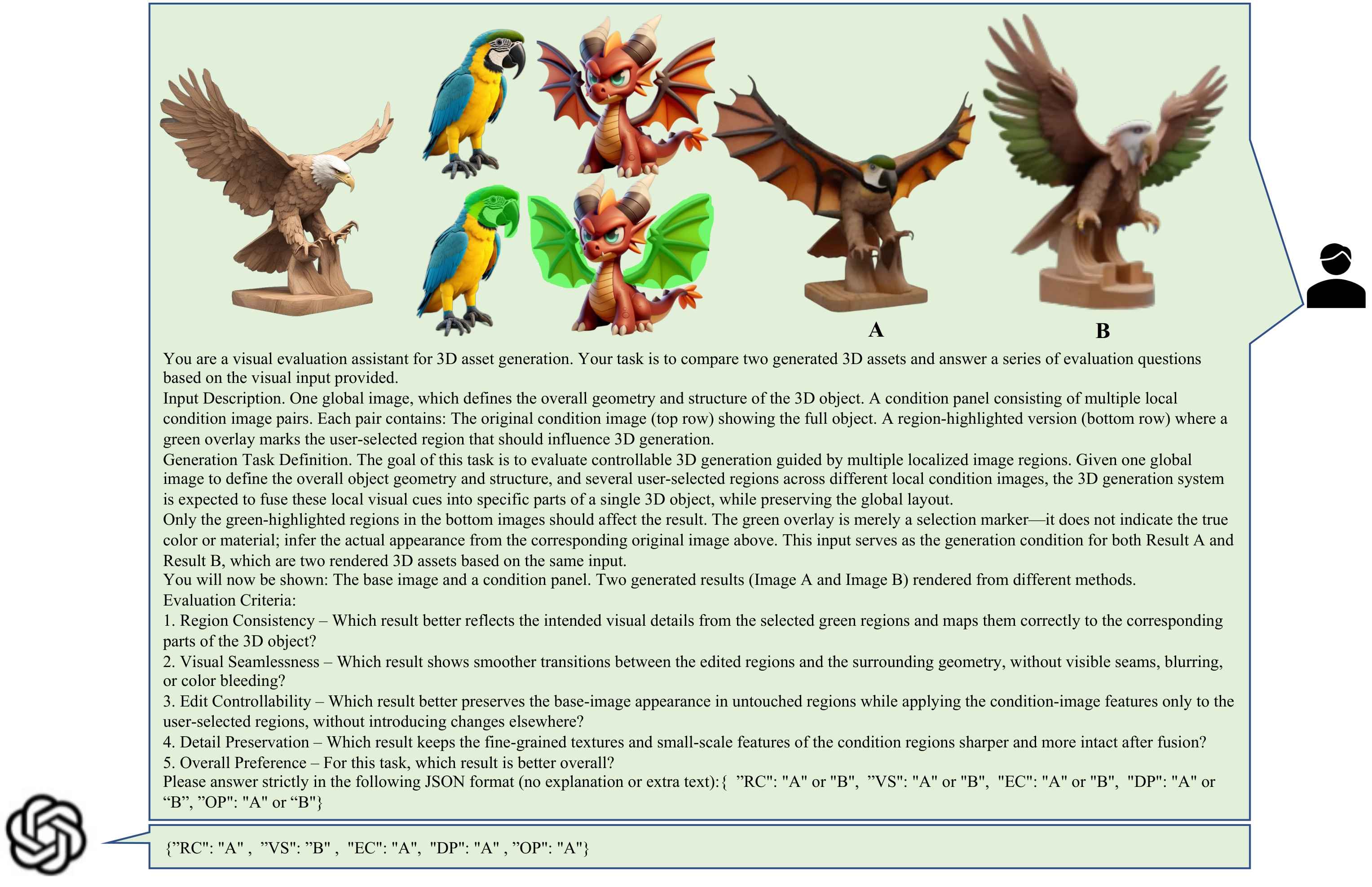}
  \caption{
    An detailed example of our GPTEval3D evaluation setup for pairwise comparison of 3D assets.
    }
  \label{fig:Supp_gptDetail}
\end{figure*}

\begin{figure}[htbp]
  \centering
  \includegraphics[width=\linewidth]{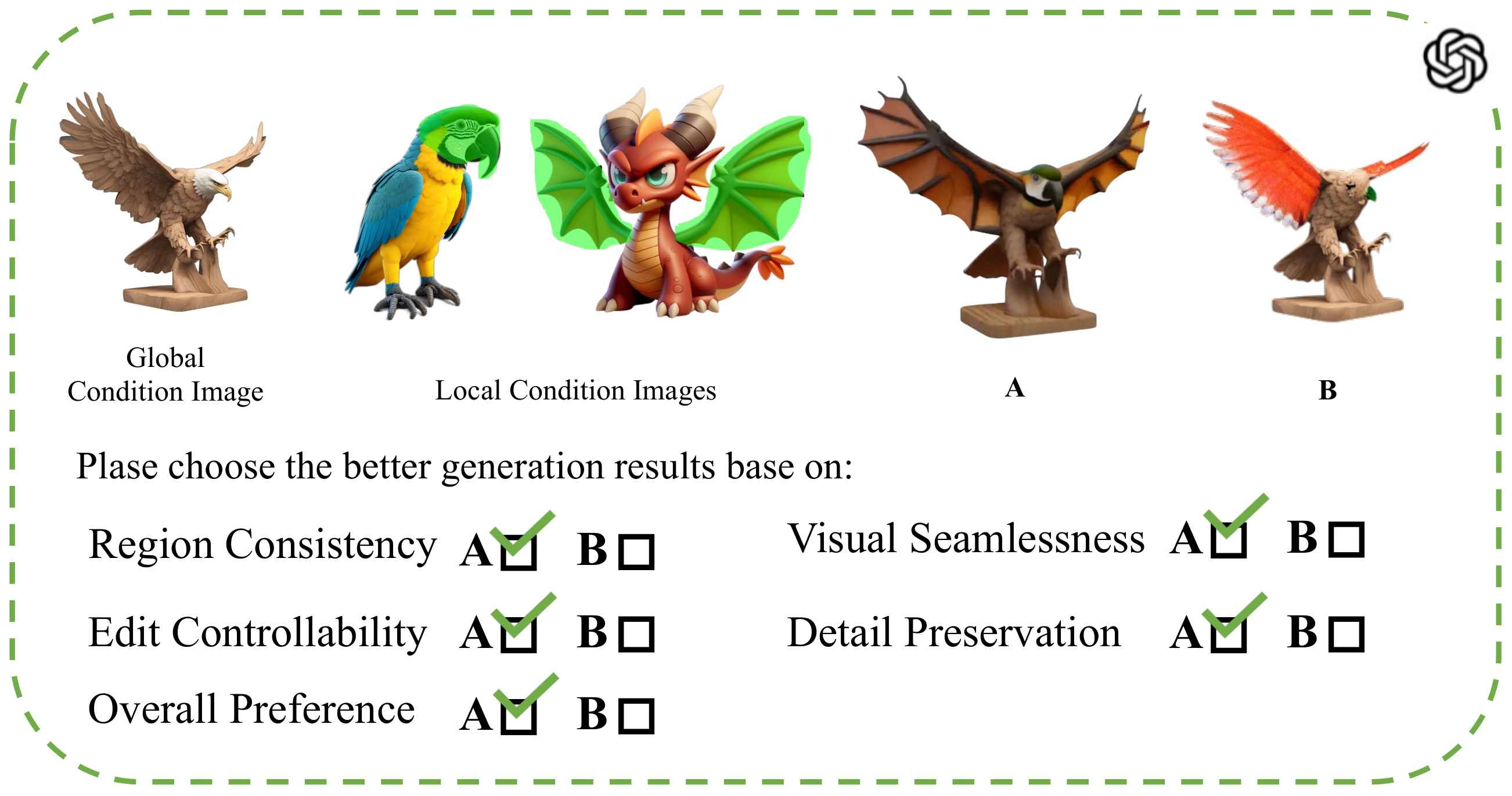}
  \caption{
    Visual illustration of the GPTEval3D evaluation process.
    }
  \label{fig:Supp_gpt}
\end{figure}
\paragraph{GPTEval3D}

While CLIP similarity and ImageReward offer useful automatic scoring for overall semantic alignment and visual appeal, they fail to capture finer-grained aspects of controllability and alignment.
To address this limitation, we adopt GPTEval3D~\cite{gptEval}, a structured evaluation framework powered by GPT-4o, which enables detailed pairwise comparisons between generation results. In this process, GPT-4 served as an objective user proxy to conduct a user study.
\refFig{fig:Supp_gpt} provides a simplified visual example illustrating how GPTEval3D is conducted.

Specifically, we prepare a set of 90 data examples, each consisting of a global condition image, multiple local condition images, and corresponding outputs generated by two different methods.
For each example, GPT-4o is provided with a fixed textual description, the input images, and two results generated by different methods, and is then prompted to decide which result performs better based on the criteria described in the textual description.
An detailed example of how GPTEval3D is performed in practice is shown in \refFig{fig:Supp_gptDetail}.
The same evaluation process is applied consistently across all examples.
The model’s responses are then aggregated to compute the final preference percentages for each method.

\subsection{Baseline Setup and Additional Baselines}

\begin{figure*}
  \centering
  \includegraphics[width=\linewidth]
  {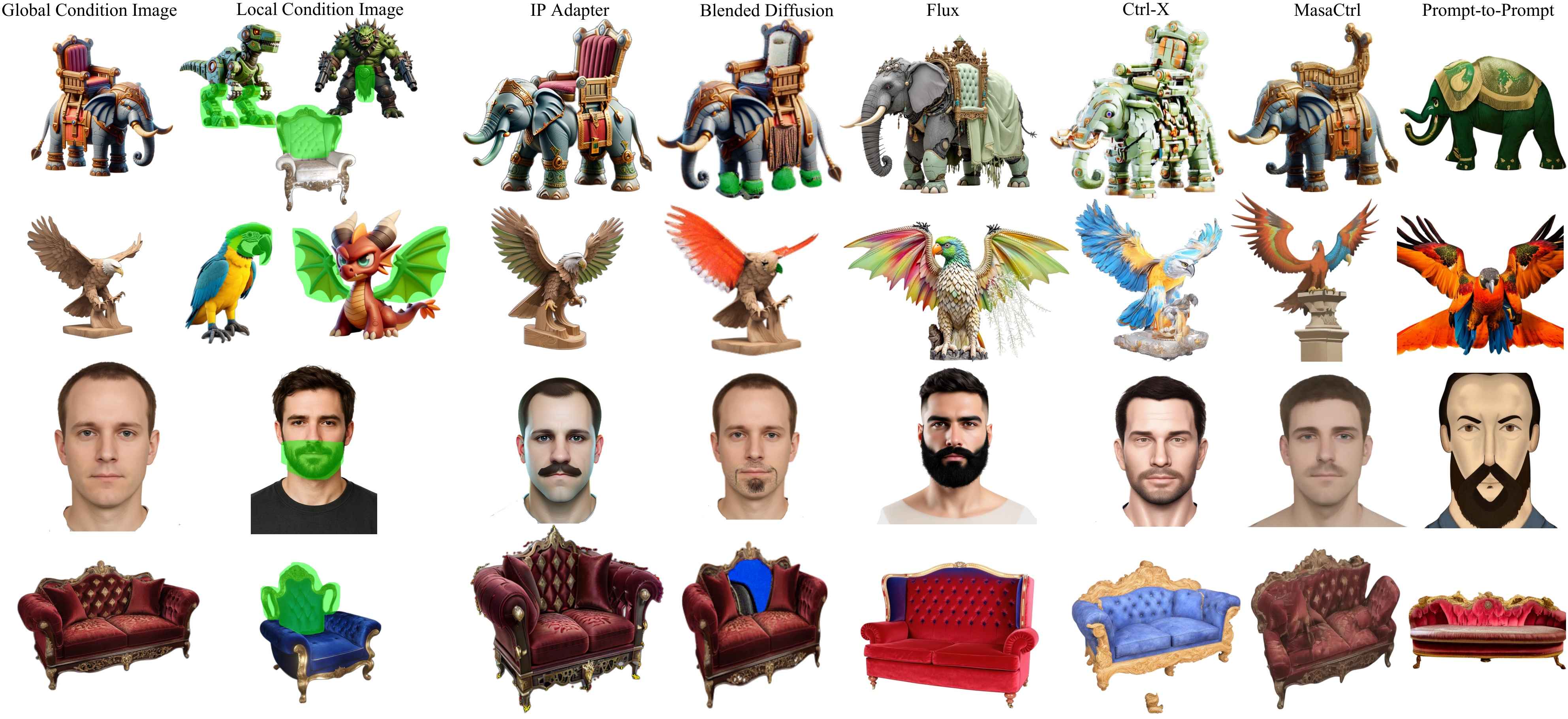}
    \caption{
    \revise{
    Intermediate fused 2D images produced by the image-based baselines. 
    These results are used as inputs to TRELLIS for 3D generation.
    }
    }
  \label{fig:intermediate_images}
\end{figure*}

\begin{figure*}
  \centering
  \includegraphics[width=\linewidth]
  {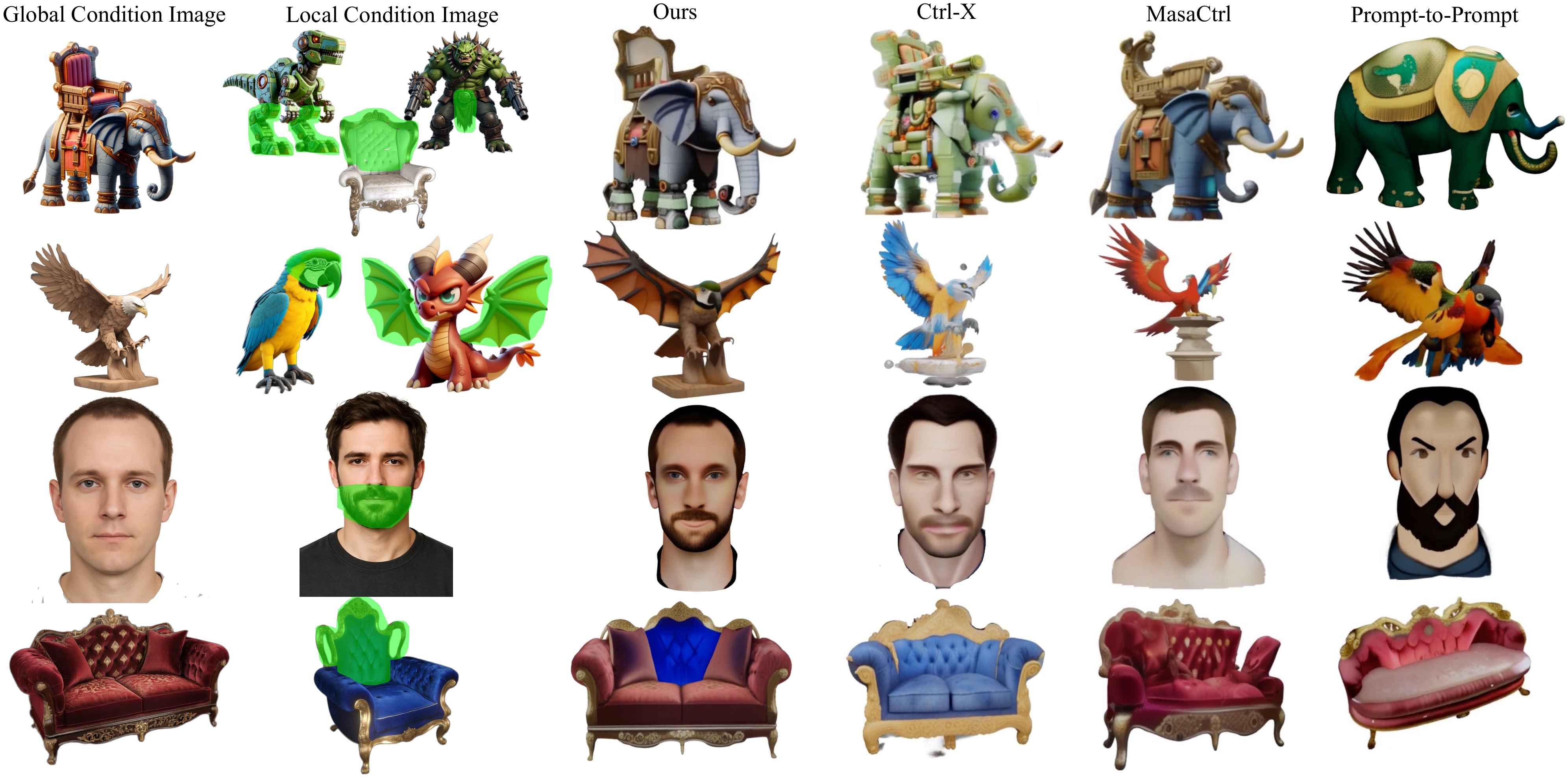}
  \caption{
  \revise{
  Comparisons with additional baselines.
  }
  }
  \label{fig:additional_baseline}
\end{figure*}

\revise{
In this subsection, we provide the detailed setup for each baseline used in the main paper, and further extend the comparison with additional baselines to ensure fairness. 
}

\revise{
\paragraph{IP-Adapter.} 
We use the official implementation of IP-Adapter~\cite{ipAdapter} with the \texttt{stable-diffusion-v1-5} checkpoint. 
Given the global condition image and the GPT\mbox{-}4o caption, IP-Adapter generates a fused 2D image. 
This image is passed into TRELLIS to produce the final 3D asset. 
}

\revise{
\paragraph{Blended Diffusion.} 
Blended Diffusion~\cite{blendedDiffusion} performs text-guided inpainting with user-defined masks, using the official checkpoint. 
We use the global condition image as the base and manually provide masks for local condition regions. 
Each region is edited with the corresponding local description to obtain a fused 2D image, which is then fed into TRELLIS. 
}

\revise{
\paragraph{FLUX.} 
We use the \texttt{flux-1.1-pro} checkpoint of FLUX~\cite{flux}, a powerful text-to-image model. 
The GPT-4o caption is provided directly as input to generate a fused image, which is then fed into TRELLIS for 3D generation. 
}

\revise{
\paragraph{TRELLIS.} 
TRELLIS~\cite{trellis} is used directly in its text-to-3D setting. 
We provide the GPT-4o caption as textual input and generate the 3D asset without any intermediate 2D fusion step. 
This baseline serves as a reference for text-only conditioning. 
}

\revise{
While Fuse3D is the first work to explicitly address multi-image, region-level fusion for 3D generation, the lack of prior art makes baseline selection debatable. 
To strengthen fairness, we additionally include three more methods as comparisons: MasaCtrl~\cite{masactrl}, Prompt-to-Prompt~\cite{prompt2prompt}, and Ctrl-X~\cite{ctrlx}. 
}

\revise{
\paragraph{MasaCtrl.}
We follow the official SDXL real-image editing pipeline of MasaCtrl~\cite{masactrl}, based on \texttt{stable-diffusion-v1-4}. 
The global condition image is used as the source, and the GPT\mbox{-}4o fused caption serves as the target prompt. 
We adopt the recommended step and layer settings from the released code, generate a fused 2D edit, and feed it into TRELLIS for 3D generation.
}

\revise{
\paragraph{Prompt-to-Prompt.} 
We adapt Prompt-to-Prompt~\cite{prompt2prompt} to our setting using the \texttt{stable-diffusion-v1-5} checkpoint. 
The GPT-4o generated source and target prompts are used as inputs, where the target prompt is aligned with the source prompt to have identical token length by padding with neutral tokens. 
The resulting fused image is then used as input to TRELLIS for 3D generation. 
}

\revise{
\paragraph{Ctrl-X.} 
We use the standard implementation of Ctrl-X~\cite{ctrlx}, built on \texttt{stable-diffusion-xl-base-1.0}. 
A global condition image and a local condition image are combined through feature fusion to produce a fused 2D image. 
This image is subsequently passed into TRELLIS to obtain the 3D asset. 
}

\revise{
The results of MasaCtrl, Prompt-to-Prompt, and Ctrl-X are presented separately for direct comparison with Fuse3D in~\refFig{fig:additional_baseline}. 
Moreover, to ensure transparency, we present the intermediate fused 2D images produced by all image-based baselines in~\refFig{fig:intermediate_images}.
}

\section{\revise{More Disscusion on Technical Design.}}

\begin{figure*}
  \centering
  \includegraphics[width=\linewidth]
  {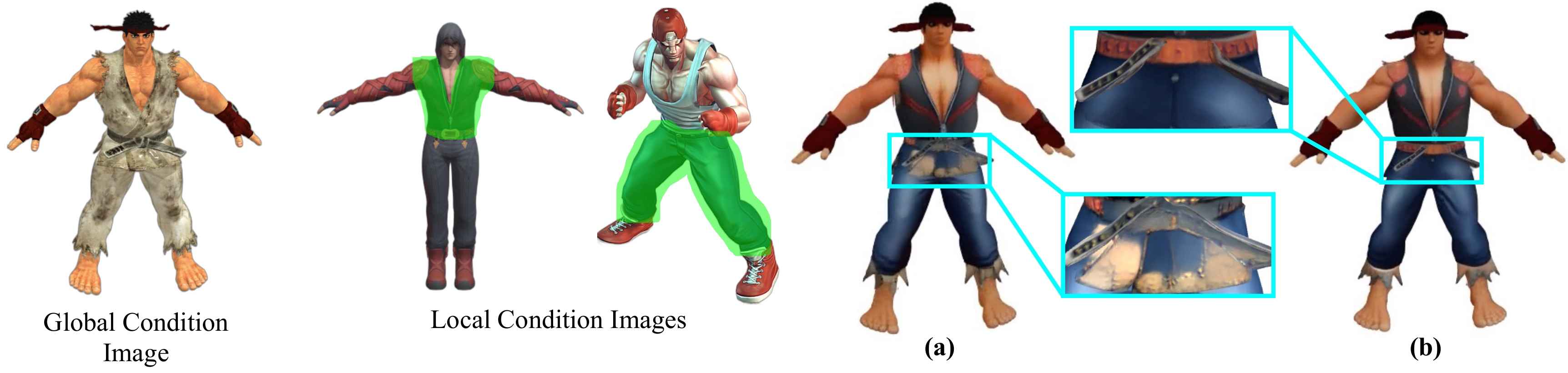}
  \caption{
  Comparison between inpainting-based strategy \textbf{(a)} and our proposed method \textbf{(b)}.
  }
  \label{fig:Supp_inpaint}
\end{figure*}

\subsection{Prepended Tokens In DINOv2}

In this section, we investigate the role of the \texttt{[CLS]} token and the \texttt{[REG]} tokens in guiding the generation process.
According to DINOv2~\cite{dinov2, dinoReg}, the \texttt{[CLS]} token functions similarly to the \texttt{[class]} token in BERT~\cite{bert} and is primarily used for classification.
The \texttt{[REG]} tokens are introduced to capture additional global contextual information.

Notably, in the original DINOv2+reg configuration, the \texttt{[REG]} tokens are discarded after the transformer layers, with only the \texttt{[CLS]} token and patch tokens used for downstream tasks.
However, the pretrained model we adopt retains both the \texttt{[CLS]} and \texttt{[REG]} tokens throughout the network, and we preserve them during the MCFM as part of the unified condition representation.

\begin{figure}
  \centering
  \includegraphics[width=\linewidth]{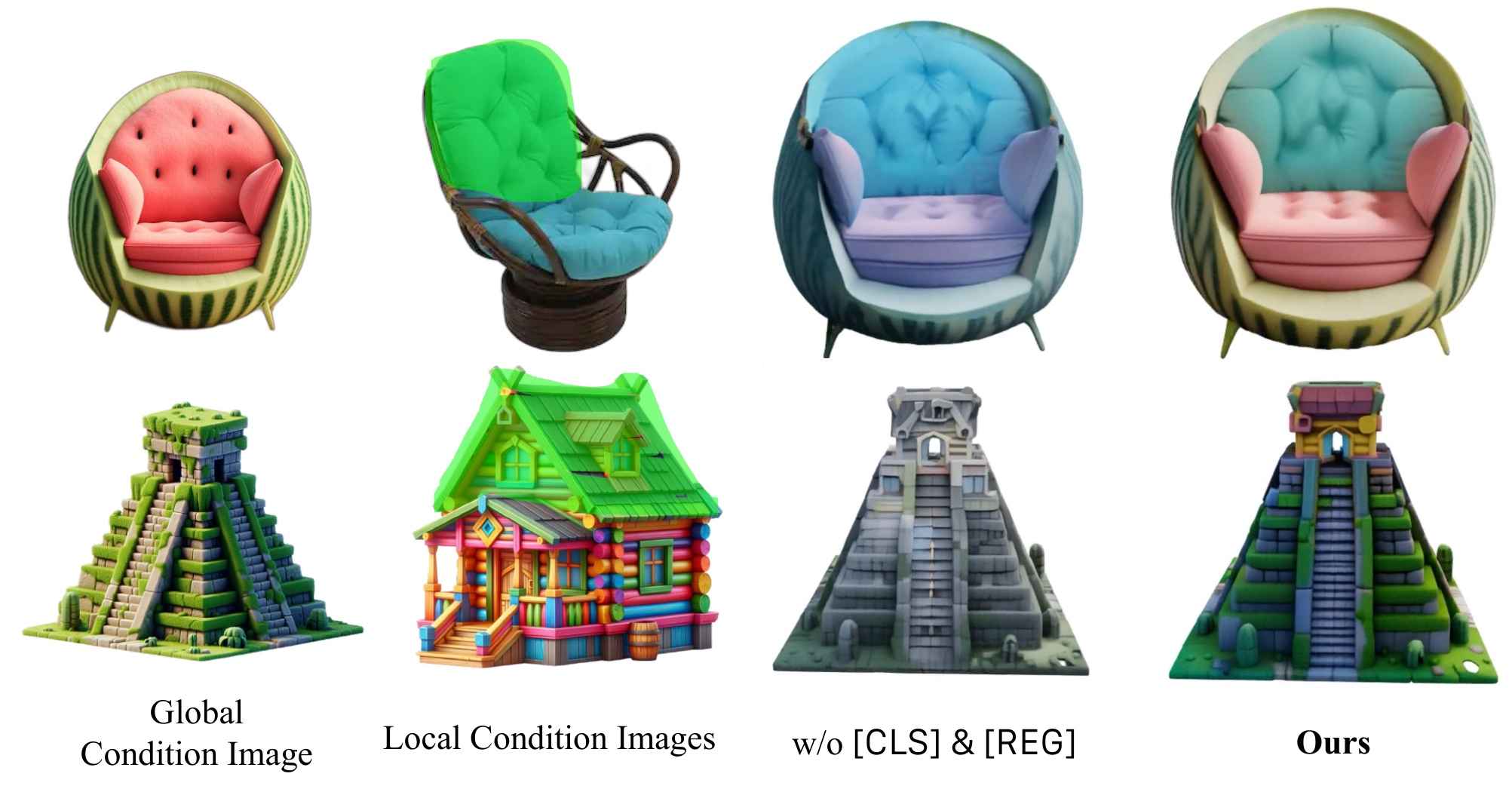}
  \caption{
   Analysis on \texttt{[CLS]} and \texttt{[REG]} tokens.
  }
  \label{fig:Supp_ablaNOREG}
\end{figure}

As shown in \refFig{fig:Supp_ablaNOREG}, although the model without \texttt{[CLS]} and \texttt{[REG]} tokens still captures the 2D-to-3D region correspondences, the overall generation quality and color fidelity significantly deteriorate.
For instance, in the watermelon chair example, the red cushion and green shell are severely biased in hue, producing unnatural blue-green tones.
In contrast, our method preserves color saturation and structural consistency more effectively, resulting in higher-quality 3D outputs.

\subsection{Inpainting Mechanism}

We also experiment with inpainting-based approaches~\cite{repaint, blendDiffusion} to perform multi-condition fusion generation. 
In this setup, we utilize the 2D-to-3D correspondences to inpaint the latent attached to the aligned voxels, while keeping the unaligned unchanged.




Specifically, we first generate a initial \textsc{SLat} guided by the global image, which contains both active voxels and their associated latent features.
Then, we use all local condition images as input to the MCFM module to obtain the unified condition tokens. 
Note that this unified tokens do not contain the tokens from the global image.
The unified tokens are then fed into $\boldsymbol{\mathcal{G}}_{\mathrm{L}}$ to generate new latent on the initial \textsc{SLat}.
At each sampling step $t$ of $\boldsymbol{\mathcal{G}}_{\mathrm{L}}$, for the latent $z_i^t$ attached to the unaligned voxels indexed by $\mathcal{V}_{\text{unaligned}}$, we replace $z_i^t$ with a noised version of $z_i^0$, denoted as $z_i^{Q_t}$.
Here, $z_i^0$ is the latent of the voxel $\boldsymbol{p}_i$ in the initial \textsc{SLat}, which is generated under the guidance of the global image, and $z_i^{Q_t}$ is calculated by the forward process of the flow model $\boldsymbol{\mathcal{G}}_{\mathrm{L}}$ based on the $z_i^0$.

Meanwhile, the latent attached to the aligned voxels is regenerated under the guidance of the unified condition tokens, enabling selective region updates during the sampling process.
As a result, at time step $t$, for each latent $z_i^t$ attached on the voxel $\boldsymbol{p}_i$, the update rule is:
\[
z_i^t =
\begin{cases}
z_i^{Q_t}, & \text{if} \quad i \in \mathcal{V}_{\text{unaligned}},
\\[6pt]
z_i^t, & \text{if} \quad i \in \bigcup_{k=1}^N \mathcal{V}_k.
\end{cases}
\]

\refFig{fig:Supp_inpaint} compares the inpainting-based strategy with our proposed approach.
We observe that the result produced by the inpainting approach exhibits noticeable holes and incomplete surface coverage.
This occurs because in certain cases, aligned voxels themselves are sparse or partially missing, and the inpainting mechanism simply reuses the initial latent features attached to those misaligned voxels.

In contrast, our method leverages both the global and local condition images as input to the MCFM to generate unified condition tokens. 
These tokens are then injected into $\boldsymbol{\mathcal{G}}_{\mathrm{L}}$ during generation, allowing the model to semantically fuse features even when the aligned voxels are incomplete.
Rather than rigidly keeping features onto potentially misaligned regions, our approach enables smoother and more coherent fusion, producing geometrically and visually consistent results.

\subsection{\revise{Limitations of Image-based Editing under Occlusion}}

\begin{figure}
  \centering
  \includegraphics[width=\linewidth]
  {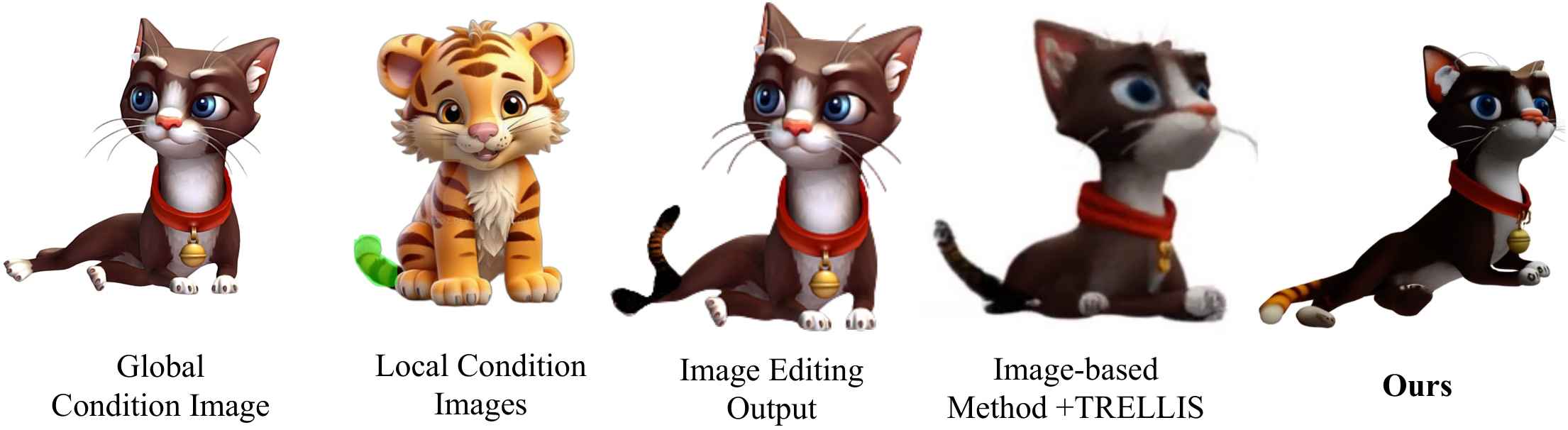}
    \caption{
    \revise{
    Comparisons with image-based methods in cases where the target 3D region is occluded in the global image.
    }
    }
  \label{fig:compare_image_based}
\end{figure}

\revise{
As shown in~\refFig{fig:compare_image_based}, we present a case where the target part is occluded in the global image (cat’s tail). A 2D image-editing pipeline (blended diffusion) cannot modify an invisible region before passing to TRELLIS, while Fuse3D transfers the tail attributes from the local condition into the correct 3D region via our 3D Semantic-Aware Alignment module. 
}

\subsection{\revise{Handling Conflicting Local Conditions}}

\begin{figure}
  \centering
  \includegraphics[width=\linewidth]
  {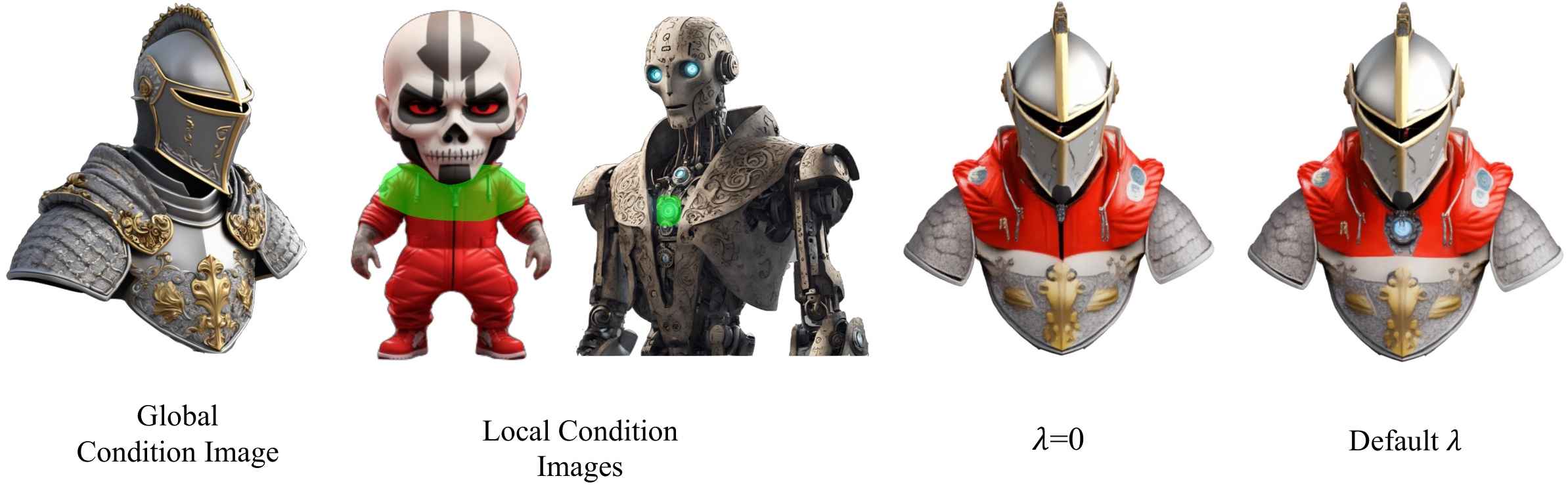}
    \caption{
    \revise{
    Cases where multiple condition images target the same 3D region.
    }
    }
  \label{fig:same_region}
\end{figure}

\revise{
We further evaluate the robustness of Fuse3D when multiple local conditions target the same 3D region with conflicting attributes in~\refFig{fig:same_region}. 
When $\lambda=0$, if multiple 2D regions target the same 3D area, it becomes uncontrollable which local condition dominates that region. 
As shown in~\refFig{fig:same_region}, the red armor attribute overrides the intended blue gem.
With our default setting, $\lambda_k$ is determined by the proportion of the selected region relative to the entire condition image. 
This design ensures that more precisely selected regions (i.e., with fewer selected tokens) exert stronger and more localized influence on the corresponding 3D voxels. 
As a result, the small region specifying the blue gem dominates over the conflicting red armor, yielding a coherent 3D result.
}

\subsection{\revise{Effect of the Number of Condition Images}}

\begin{table}[!htp]
\centering
\caption{
\revise{
Region-specific CLIP similarity with different numbers of condition images. 
}
}
\revise{
\begin{tabular}{lccc}
\toprule
Number of Condition Images & 1 & 2 & 3  \\
\midrule
Region-specific CLIP $\uparrow$ & 0.242 & 0.239 & 0.240 \\
\bottomrule
\end{tabular}
}
\label{tab:num_images}
\end{table}

\revise{
We also examine how the number of condition images affects the generation quality. 
For this experiment, we varied the number of local condition images from 1 to 3 and computed the region-specific CLIP similarity between each fused region and its source. 
This evaluation measures how consistently Fuse3D preserves local attributes under different input settings.
}

\revise{
As shown in~\refTab{tab:num_images}, the CLIP similarity scores remain stable 
across different numbers of condition images. 
This indicates that Fuse3D is robust to the number of inputs and can maintain faithful region-level alignment regardless of how many condition images are provided. 
The results further suggest that users are free to supply one or multiple condition images without significantly impacting performance.
}

\subsection{\revise{User Study on the Enhancement Factor $\lambda$}}

\begin{table}[!htp]
\centering
\caption{
\revise{
User ratings (1--5) on Source Separation under different enhancement factor $\lambda$. 
}
}
\revise{
\begin{tabular}{lcccc}
\toprule
$\lambda$ & 0.5 & 1.0 & 2.0 & 5.0 \\
\midrule
Source Separation $\uparrow$ & 3.7 & 3.8 & 3.8 & 4.2 \\
\bottomrule
\end{tabular}
}
\label{tab:user_lambda}
\end{table}

\revise{
To evaluate how the local attention enhancement factor $\lambda$ influences the final results, 
we conducted a user study focusing on whether different local images remain clearly separated in their corresponding 3D regions, or whether they undesirably mix together. 
Participants were asked to rate the clarity of source separation on a 1--5 Likert scale for results generated with $\lambda \in \{0.5, 1.0, 2.0, 5.0\}$. 
A higher score indicates cleaner separation with less fusion across regions.
}

\revise{
The averaged scores are reported in~\refTab{tab:user_lambda}. 
We observe that increasing $\lambda$ generally strengthens the separation of local conditions. 
The results demonstrate that the enhancement factor effectively adjusts the fusion strength of local control features, enabling a balance between local separation and global coherence.
}

\subsection{More Ablation Experiment}

\begin{figure}
  \centering
  \includegraphics[width=\linewidth]{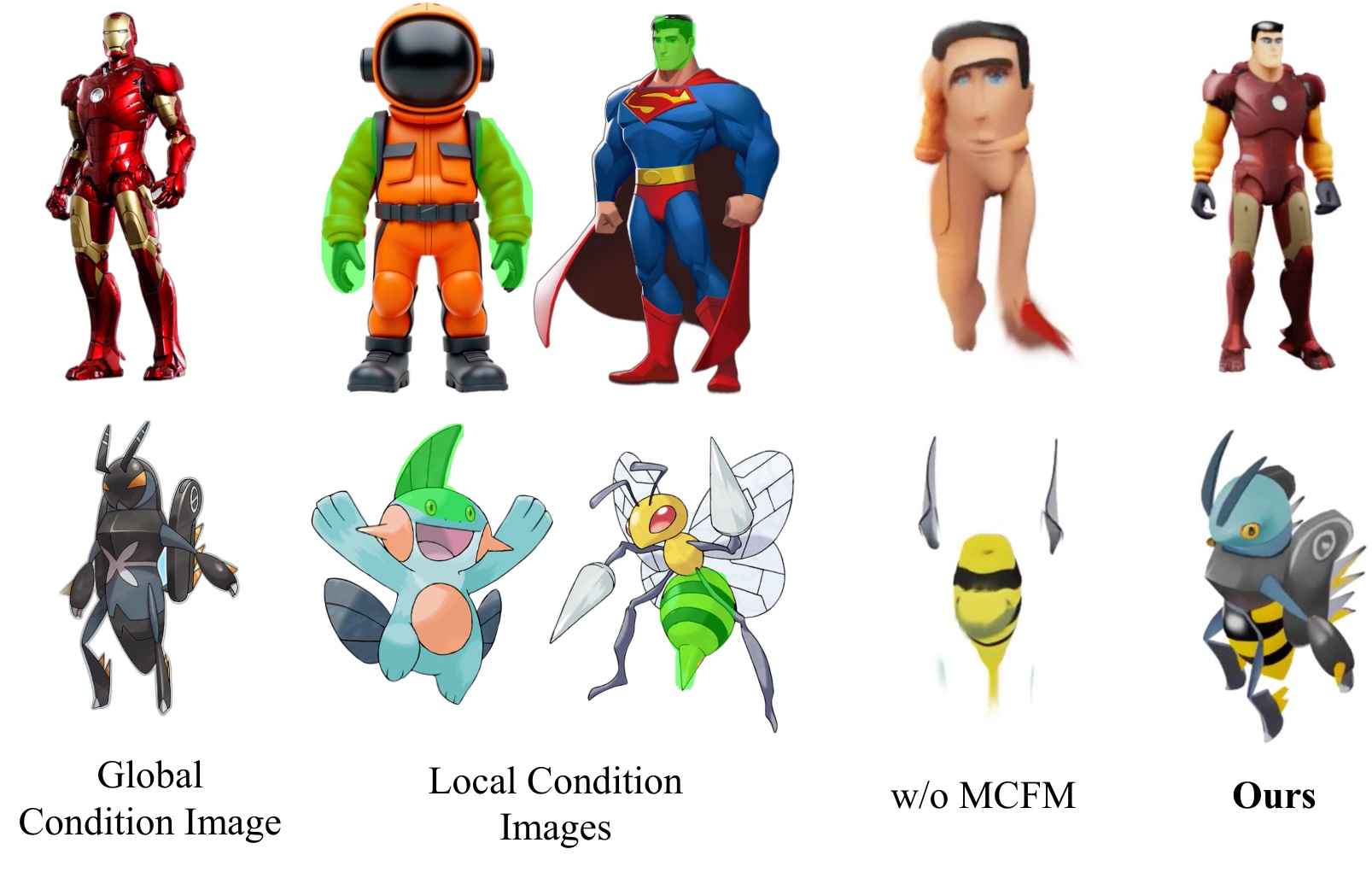}
  \caption{
  More ablation analysis of MCFM.
  }
  \label{fig:Supp_ablaMCFM}
\end{figure}

\begin{figure}
  \centering
  \includegraphics[width=\linewidth]{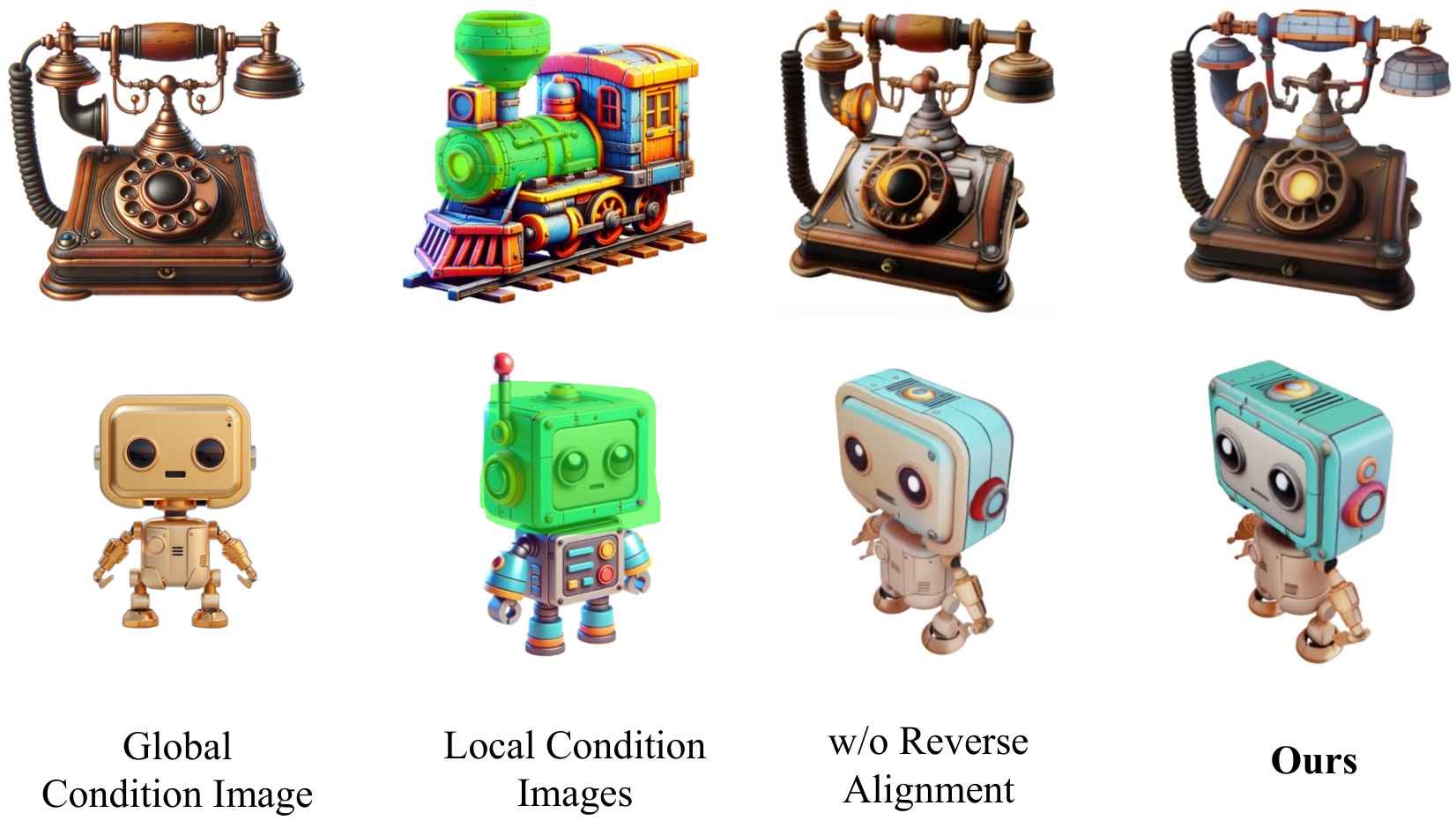}
  \caption{
  More ablation analysis of the reverse alignment component in the 3D Semantic-Aware Alignment Module.
  }
  \label{fig:Supp_ablaRA}
\end{figure}

\begin{figure}
  \centering
  \includegraphics[width=\linewidth]{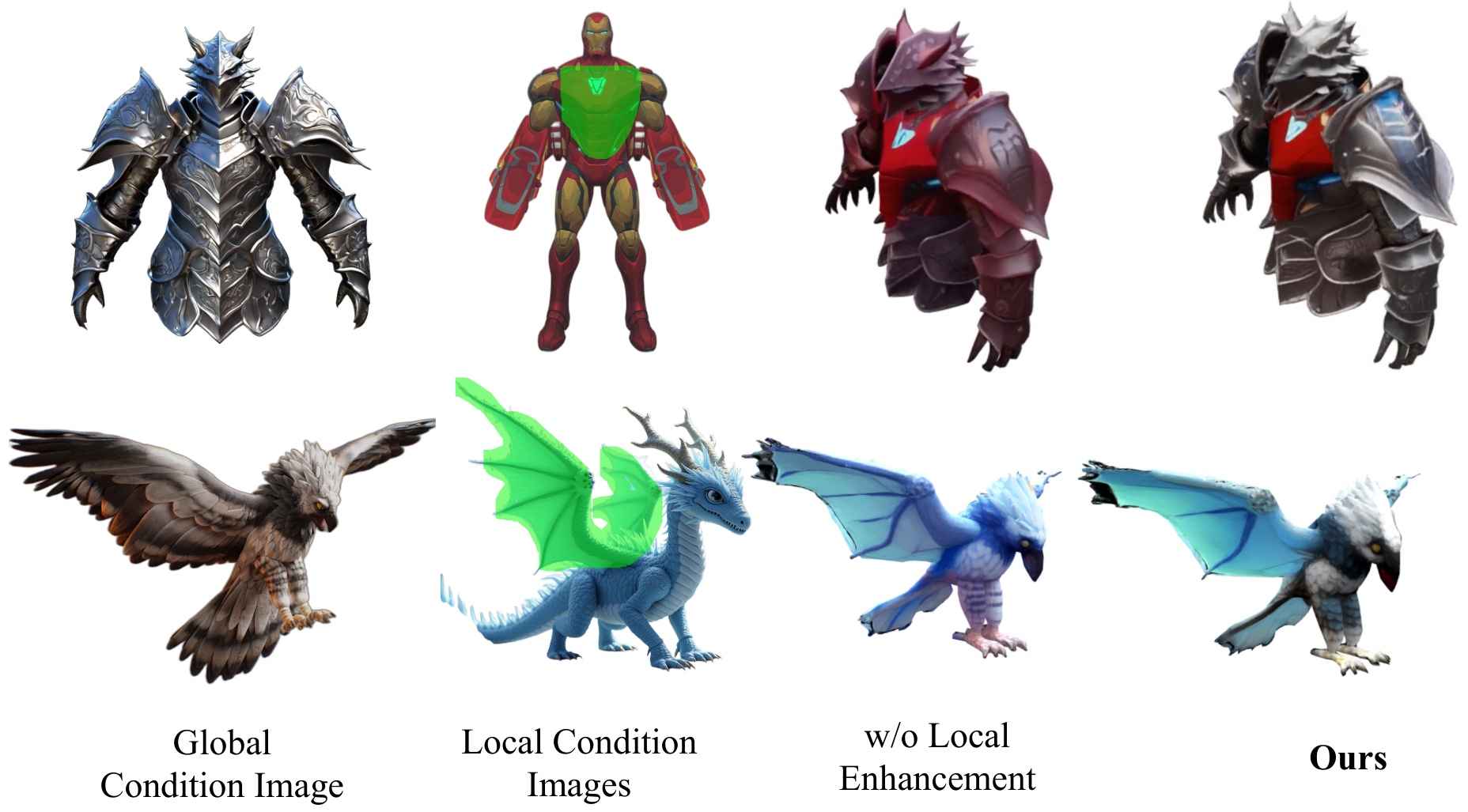}
  \caption{
  More ablation analysis of local attention enhancement module.
  }
  \label{fig:Supp_ablaLocalEnhance}
\end{figure}

We present more ablation experiment here.

\paragraph{MCFM}

As shown in \refFig{fig:Supp_ablaMCFM}, removing the MCFM module leads to severe degradation in structure and spatial coherence.
In the ablation study setup, we crop each selected local patch in the image space and feed it into the encoder to obtain the image tokens corresponding to the cropped patches. The tokens from multiple cropped images are concatenated together to guide the subsequent 3D generation.
This approach breaks the spatial context that DINOv2 relies on, leading to incorrect part placement and geometric distortions
This issue is most clearly seen in the Superman example. The model maps the face directly onto the torso, indicating that it lacks the global understanding needed to place the face at the correct head position.

By contrast, our method incorporates MCFM to retain global prior information while integrating multiple local features. This enables the fused results to maintain correct part alignment, overall geometry, and consistent feature transfer across the entire object.

\paragraph{Reverse Alignment Module}
Since completely removing the 3D Semantic-Aware Alignment would prevent subsequent local enhancement, we retained the Forward Alignment while removing the Reverse Alignment Module in the ablation study to compare the results.
As shown in \refFig{fig:Supp_ablaRA}, removing the reverse alignment component leads to unintended overlap between features from global condition image and local condition images.
In the robot example, the fused head is influenced by both blue and golden robots, creating a visually inconsistent and structurally ambiguous result.

This occurs because, without reverse alignment, the model fails to filter out the tokens that target the same voxel, resulting in unintended feature fusion.
Consequently, features from both global and local images are fused in the same region, leading to visual ambiguity.

\paragraph{Local Attention Enhancement Module}

The local attention enhancement module is designed to regulate the fusion when features from different images affect nearby voxels. Without this module, features from different condition images tend to conflict and interfere with each other during the generation process.

As shown in the Ironman example in \refFig{fig:Supp_ablaLocalEnhance}, the red color from the Ironman image is overflows and gets integrated into the silver-armored character, resulting in a blended appearance.
In contrast, with the local attention enhancement module, the features from different images remain more distinctly separated, yielding significant local details.

\section{Limitations}
\begin{figure}[!htp]
    \centering
    \includegraphics[width=\linewidth]{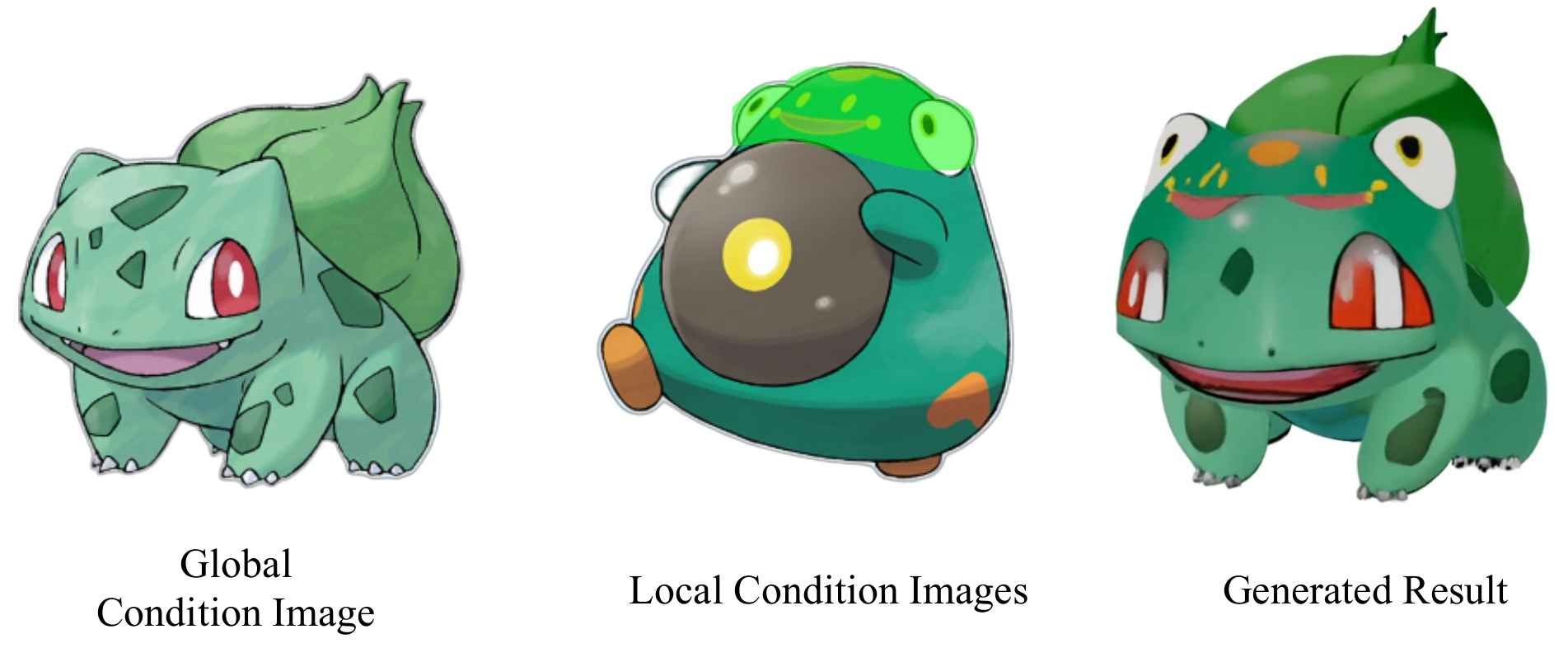}
    \caption{
        An example of our failure case where the selected 2D region are projected onto wrong 3D regions.
    }
   \label{fig:Supp_fail}
\end{figure}

While our method demonstrates strong performance in generating controllable 3D assets, it still has several limitations.
First, \revise{as shown in~\refFig{fig:Supp_fail}, the automatic 3D semantic-aware alignment module may generate inaccurate 2D-to-3D correspondences,} particularly when the feature of the selected 2D regions from multiple condition images deviate significantly.
Second, since our model is built upon the TRELLIS framework, it inherits its limitations.
For instance, our generated results may exhibit baked-in shading or highlights directly carried over from both global and local images. \revise{However, the insights proposed by our method are applicable to other 3D-native generation models, enabling seamless integration with future advancements in more powerful 3D generation models.}

\end{document}